\documentclass[aps,prl,twocolumn,superscriptaddress,showpacs,longbibliography]{revtex4-2}
\usepackage{braket}
\usepackage{graphicx}
\usepackage{latexsym}
\usepackage{amssymb}
\usepackage{amsmath}
\usepackage{amsfonts}
\usepackage{bm}
\usepackage{multirow}
\usepackage{color}
\usepackage{xcolor}
\usepackage{calrsfs}
\usepackage{dutchcal}

\usepackage{titlesec}
\titlespacing*{\section}{0pt}{5pt}{5pt}

\usepackage[colorlinks=true, citecolor={blue!80!black}, urlcolor={blue!50!black}, linkcolor = {blue!80!black},hypertexnames=false]{hyperref}

\usepackage[percent]{overpic}
\usepackage{tabularx}

\newcommand{\be}{\begin{equation}}
\newcommand{\ee}{\end{equation}}
\newcommand{\bea}{\begin{eqnarray}}
\newcommand{\eea}{\end{eqnarray}}

\begin{document}


\title{Magnetization Plateaux as a Roadmap to Quantum Spin Liquids}

\author{Anna Keselman}
\affiliation{Department of Physics, Technion, Haifa, 3200003, Israel} 
\author{Xinyuan Xu}
\affiliation{Department of Physics and Astronomy, University of Utah,  Salt Lake City, Utah 84112, USA}
\author{Hao Zhang}
\affiliation{Department of Physics and Astronomy, University of Tennessee, Knoxville, TN 37996, USA}
\author{Cristian D. Batista}
\affiliation{Department of Physics and Astronomy, University of Tennessee, Knoxville, TN 37996, USA}
\affiliation{Neutron Scattering Division, Oak Ridge National Laboratory, Oak Ridge, Tennessee 37831, USA}
\author{Oleg A. Starykh}
\affiliation{Department of Physics and Astronomy, University of Utah,  Salt Lake City, Utah 84112, USA}

\begin{abstract}
We investigate the spin-$\tfrac{1}{2}$ $J_{1}$--$J_{2}$ triangular-lattice Heisenberg antiferromagnet in a magnetic field by combining large-scale density matrix renormalization group (DMRG) simulations with self-consistent spin-wave theory. The resulting field--coupling phase diagram reveals that quantum fluctuations stabilize coplanar order across the entire parameter range, giving rise to a characteristic sequence of magnetization plateaux. Near the quantum-spin-liquid window $0.06 \!\lesssim\! J_{2}/J_{1} \!\lesssim\! 0.16$, which extends to magnetic field $B \sim J_1$, we identify overlapping $m\!=\!1/3$ and $m\!=\!1/2$ plateaux---a distinctive hallmark of the system’s proximity to the low-field spin-liquid regime. The excellent quantitative agreement between DMRG and self-consistent one-loop spin-wave calculations demonstrates that semiclassical approaches can reliably capture and parameterize the plateau phases of triangular quantum antiferromagnets. 
\end{abstract}

\maketitle


\emph{Introduction. - } 
The spin-$1/2$ triangular lattice antiferromagnet has played a central role in the search for quantum spin liquids (QSLs) since Anderson’s seminal proposal of a resonating valence bond state in 1973~\cite{Anderson1973}. Although the nearest-neighbor model ultimately develops magnetic order, extensive numerical studies have established a QSL ground state in the extended $J_1$-$J_2$ triangular lattice antiferromagnet with antiferromagnetic next-nearest-neighbor exchange $J_2$~\cite{Imada2014,Iqbal2016,White2015,Hu2015,Gong2017,Saadamand2017,He2019,Gallegos2025,Wietek2024,Bishop2015,Oitmaa2020}. Variational Monte Carlo, DMRG, exact diagonalization, coupled-cluster, and series-expansion approaches consistently find the absence of long-range magnetic order in the interval $J_2/J_1 \in (0.06,0.16)$. Furthermore, a growing body of evidence favors a $U(1)$ Dirac QSL in this regime based on comparisons of variational energies and the quantum numbers of low-energy excitations~\cite{Iqbal2016,Gong2017,Saadamand2017,Gallegos2025,Hermele2005,Song2019,He2019,Wietek2024,Sasank2025}.

This theoretical progress has galvanized the search for quantum materials realizing the target QSL regime of the triangular-lattice $J_1$-$J_2$ antiferromagnet. Prominent candidates include $A$YbSe$_2$ ($A=$Cs, K, Na)~\cite{Tsirlin_2020,Scheie2024} and YbZn$_2$GaO$_5$~\cite{Haravifard2024}, which host Yb$^{3+}$ pseudo-spin-$1/2$ moments on nearly isotropic triangular lattices with weak exchange anisotropies. Determining $J_2/J_1$, however, remains challenging. For $A$YbSe$_2$, nonlinear spin-wave fits to the spin-excitation spectra in the $m=1/3$ plateau yield $J_2/J_1 \approx 0.03$, $0.04$, and $0.07$ for $A=$Cs~\cite{Xie2023}, K, and Na~\cite{Scheie2024}, respectively. For YbZn$_2$GaO$_5$, estimates rely on comparisons between the dynamical structure factor and DMRG simulations~\cite{Haravifard2024}, which remain difficult to quantify with high confidence.

\begin{figure}
    \centering
    \includegraphics[width=\linewidth]{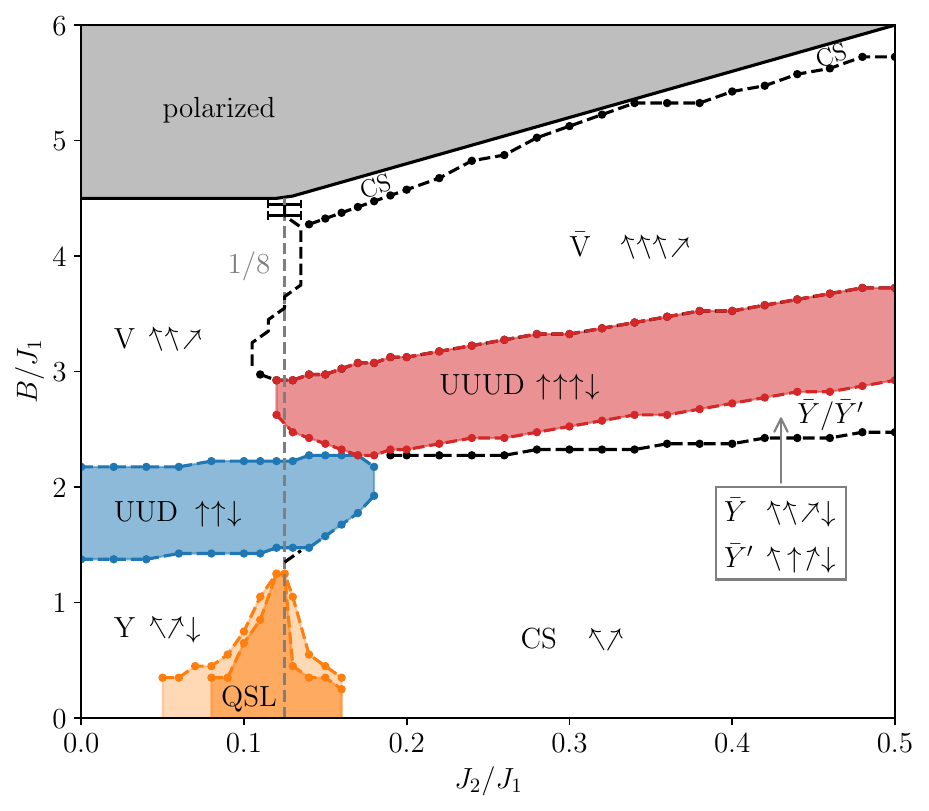}
    \caption{Phase diagram of the model obtained using 
    iDMRG on cylinders. In the low $J_2$ regime three-sublattice ordered states, Y, UUD, and V, are observed. In the large $J_2$ regime two-sublattice canted-stripes (CS) state is observed at low and at high fields. At intermediate fields four-sublattice ordered states, $\bar{Y}$/$\bar{Y}'$, UUUD, and $\bar{V}$ are observed. The arrows depict the spin configuration on each sublattice (all states are coplanar). 
    A quantum disordered regime is observed in the vicinity of $J_2/J_1=1/8$ (dashed gray line) extending up to fields $B\sim J_1$. 
    }
    \label{fig:PhaseDiagram}
\end{figure}

Our central idea is to analyze the system’s response to  magnetic field across the phase diagram  of the $S=1/2$ Hamiltonian
\begin{equation}
H = J_1 \!\sum_{\langle i,j\rangle} \mathbf{S}_i\!\cdot\!\mathbf{S}_j 
  + J_2 \!\sum_{\langle\langle i,j\rangle\rangle} \mathbf{S}_i\!\cdot\!\mathbf{S}_j
  - B\!\sum_i S_i^z .
\end{equation}
with $J_1>0$ and $J_2>0$.
For fixed $J_2/J_1$, we track the sequence of field-driven phases from zero field to saturation. In spin-1/2 materials, the required saturation fields, $B_{\mathrm{sat}}=4.5J_1$ for $J_2/J_1\leq1/8$ and $B_{\mathrm{sat}}=4 (J_1+J_2)$ for $J_2/J_1\geq 1/8$, correspond to only a few tesla for gyromagnetic factors which are not much smaller than one and are readily accessible experimentally; indeed, several of the cited works have already measured field-dependent excitation spectra~\cite{Scheie2024,Xie2023}. By mapping the $J_2/J_1$--$B$ phase diagram, we uncover a distinctive signature of the QSL regime: the emergence of \emph{two} overlapping magnetization plateaux at $m=1/3$ and $m=1/2$ (Fig.~\ref{fig:PhaseDiagram}). Their simultaneous presence provides a clear indicator of a material’s location along the $J_2/J_1$ axis. We also present results on the field evolution of the QSL for $B\!\lesssim\!J_1$.

\emph{Summary of semiclassical results.—}
Classically, the TLAFM exhibits an extensive \emph{accidental} degeneracy: the Hamiltonian decomposes into sums of squared three-spin (for $J_2/J_1<1/8$) or four-spin (for $J_2/J_1>1/8$) units, yielding infinitely many degenerate ground states. In the semiclassical regime, quantum zero-point fluctuations lift this degeneracy and select coplanar $120^\circ$ order for $J_2/J_1<1/8$ and collinear stripe order for $J_2/J_1>1/8$~\cite{Jolicoeur1990,Chubukov1992,Korshynov1993}.

A striking feature is that the preference for \emph{coplanar} order persists under an applied magnetic field: quantum fluctuations select coplanar states for all $J_2\le J_1$. For $J_2/J_1<1/8$, this yields the familiar three-sublattice Y, UUD (the $m=1/3$ plateau), and V states~\cite{Chubukov_1991,Starykh_2015}. For $J_2/J_1>1/8$, the two-sublattice coplanar stripe phase brackets the four-sublattice analogues $\bar{V}$, UUUD (the $m=1/2$ plateau), and $\bar{Y}$ states, as established by Chubukov and Ye~\cite{Ye2017,Ye2017b}. Classically, these three- and two/four-sublattice families meet at a field-independent first-order transition at $J_2=1/8$. Throughout this phase diagram, the scalar chirality remains zero.

\emph{The phase diagram. - } We have performed extensive simulations of the model using infinite DMRG (iDMRG) on cylinders employing the TeNPy library~\cite{tenpy2024}.
We denote by XC(YC)-$N_y$ cylinders with $N_y$ sites along the circumference in which one of the triangular lattice basis vectors is parallel (perpendicular) to the cylinder axis.
Simulations were performed mainly on XC-8 cylinders with unit-cell length $N_x=12$.
Our findings are summarized in the phase diagram in Fig. \ref{fig:PhaseDiagram}. 
Because the field couples to the conserved total magnetization $S^z_{\rm tot}=\sum_i S^z_i$, we exploit this symmetry by first obtaining the zero-field ground state in each $S^z_{\rm tot}$ sector. The magnetization at finite field is then determined by selecting, for each $B$, the sector that minimizes the total energy per site.
Bond dimensions of up to $2000$ were used, resulting in truncation errors of order $10^{-8}$ in the gapped plateaux phases, $10^{-6}-10^{-7}$ in the gapless ordered phases, and $10^{-5}$ in the QSL regime.

To identify the phases, we analyze local spin expectation values and correlation patterns. Overall, we find good qualitative agreement with the semiclassical picture, including the expected coplanar states. Notably, just below the $m=1/2$ UUUD plateau—where semiclassics predicts the $\bar{Y}$ phase—we observe a competition between $\bar{Y}$ and a distinct Y-like state, which we label $\bar{Y}'$. Both are four-sublattice coplanar states with one spin pointing down, but their tilt patterns differ: in $\bar{Y}$, the three up-spins tilt such that two lean in one direction and the third in the opposite, whereas in $\bar{Y}'$ one up-spin remains untilted and the other two tilt oppositely. These phases can be distinguished unambiguously by the characteristic peak patterns of their static structure factors, discussed below. In addition, just below saturation, we observe a competition between the canted stripe phase and a supersolid stripe phase that exhibits modulations in the longitudinal component of the ordered moment.

To locate phase boundaries, we compute the spin structure factor
\begin{equation}
    S({\bf q})=\frac{1}{N}\sum_{{\bf r},{\bf r}'}e^{-i{\bf q}\cdot({\bf r}- {\bf r}')}\langle {\bf S}_{\bf r} \cdot {\bf S}_{{\bf r}'}\rangle,
\end{equation}
where the sum runs over the $N$ sites of the iDMRG unit cell. Different ordered phases exhibit characteristic peak patterns in $S(\mathbf q)$. Three-sublattice states (Y, UUD, V) produce peaks at the $K$ and $K'$ points, while two- and four-sublattice states peak at the $M$ points. The $C_3$-symmetric UUUD and $\bar V$ phases show equal-intensity peaks at all three $M$ points, whereas the canted stripe phase yields a single dominant $M$-point peak due to broken $C_3$ symmetry. The $\bar Y$ and $\bar Y'$ states are best distinguished through their transverse spin correlations (see End Matter).

\begin{figure}
    \centering
    \includegraphics[trim={3cm 0 0 0},clip,width=0.48\linewidth]{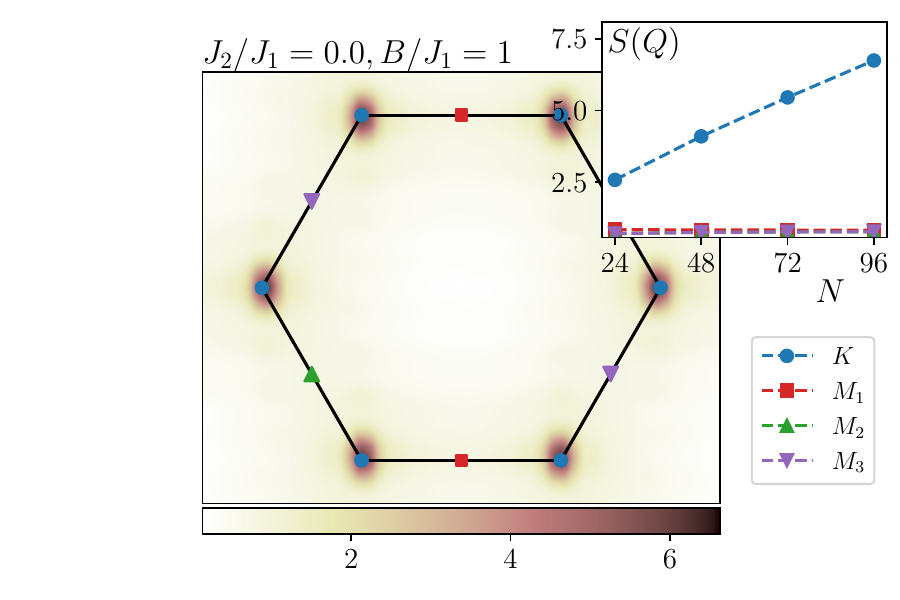}
    \llap{
    \parbox{0pt}{\vspace{-195pt}\hspace{-220pt}\footnotesize{(a)}} 
    \parbox{0pt}{\vspace{-40pt}\hspace{-210pt}\footnotesize{Y}}
    }
    \includegraphics[trim={3cm 0 0 0},clip,width=0.48\linewidth]{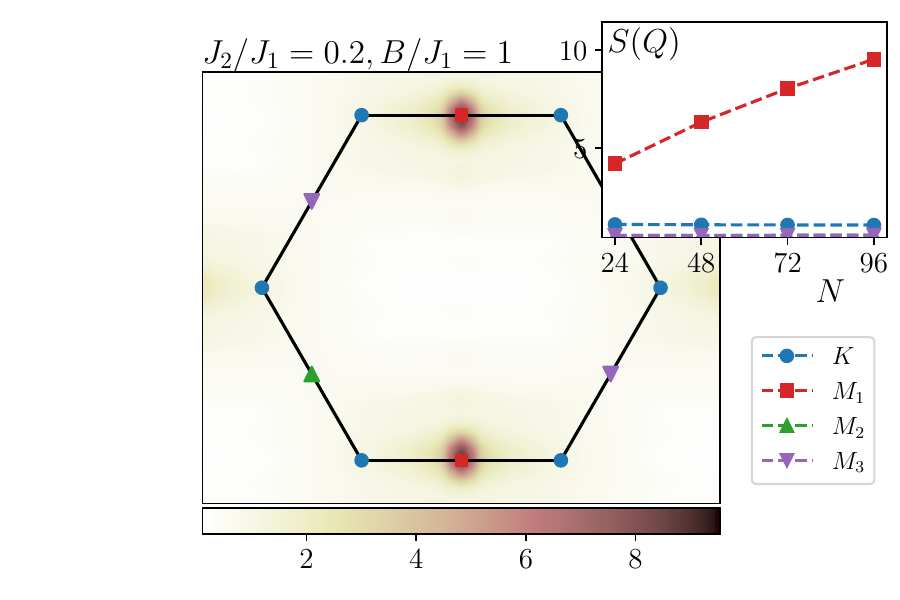}
    \llap{
    \parbox{0pt}{\vspace{-195pt}\hspace{-220pt}\footnotesize{(b)}}
    \parbox{0pt}{\vspace{-40pt}\hspace{-210pt}\footnotesize{CS}}
    }
    
    \includegraphics[trim={3cm 0 0 0},clip,width=0.48\linewidth]
    {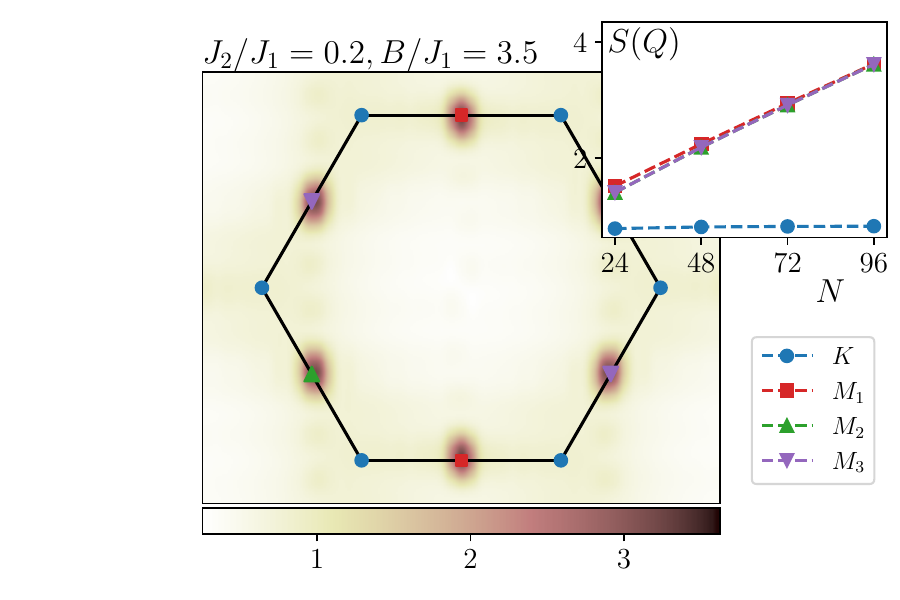}
    \llap{
    \parbox{0pt}{\vspace{-195pt}\hspace{-220pt}\footnotesize{(c)}}
    \parbox{0pt}{\vspace{-40pt}\hspace{-210pt}\footnotesize{$\bar{\rm V}$}}
    }
    \includegraphics[trim={3cm 0 0 0},clip,width=0.48\linewidth]
    {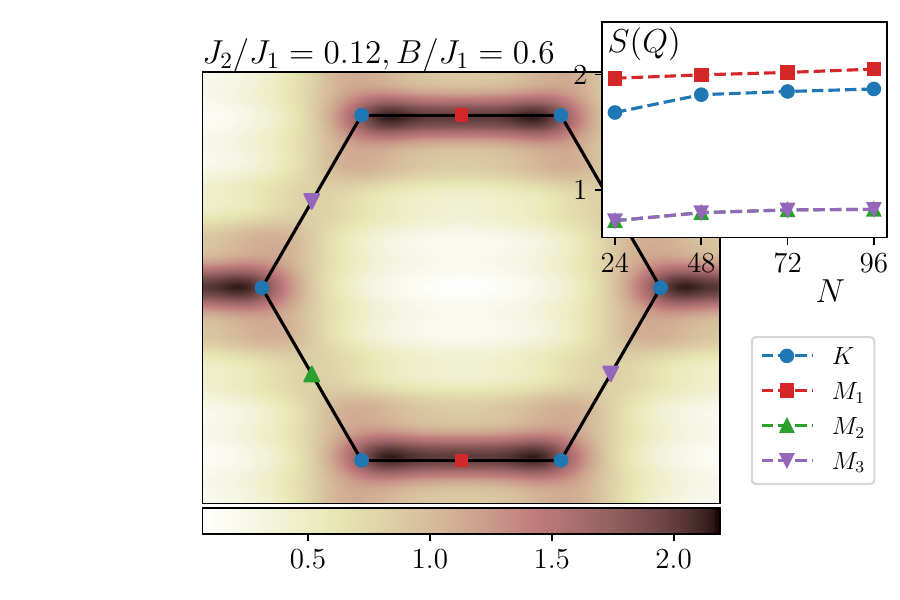}
    \llap{
    \parbox{0pt}{\vspace{-195pt}\hspace{-230pt}\footnotesize{(d)}}
    \parbox{0pt}{\vspace{-40pt}\hspace{-205pt}\footnotesize{QSL}}
    }
    \caption{Structure factor $S({\bf q})-(S^z_{\rm tot})^2/N$ calculated on XC-8 cylinders in the representative (a) three-sublattice Y phase, (b) two-sublattice canted-stripe (CS) phase (c) four-sublattice $\bar{\rm V}$ phase (d) QSL regime. Insets show  scaling of the structure factor with number of sites, $N$, at the high-symmetry points. }
    \label{fig:Sq}
\end{figure}

Representative structure factors for the various phases are shown in Fig.~\ref{fig:Sq}, where we plot $S(\mathbf q)-(S^z_{\rm tot})^2/N$ to remove the trivial $\Gamma$-point contribution. 
To identify the ordering wave vectors and thus the phase boundaries between the ordered phases more carefully, we analyze the scaling of $S(\mathbf q)$ at high-symmetry points with the number of sites $N$, as illustrated in the insets. As expected, $S(\mathbf Q)$ grows rapidly in the ordered phases, while remaining nearly constant in the QSL regime. 
Magnetization plateaux are identified as three- and four-sublattice states with magnetization fixed at $1/3$ and $1/2$ of saturation, respectively.

Below, we focus on different regimes in the phase diagram addressing and further characterizing the phases and phase transitions, with numerical results supported by spin-wave calculations.

\begin{figure}
    \centering
    \includegraphics[width=0.48\linewidth]{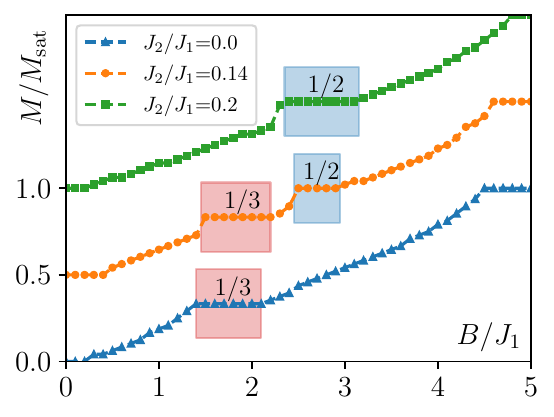}
    \llap{\parbox{0pt}{\vspace{-180pt}\hspace{-230pt}\footnotesize{(a)}}} 
    \includegraphics[width=0.48\linewidth]{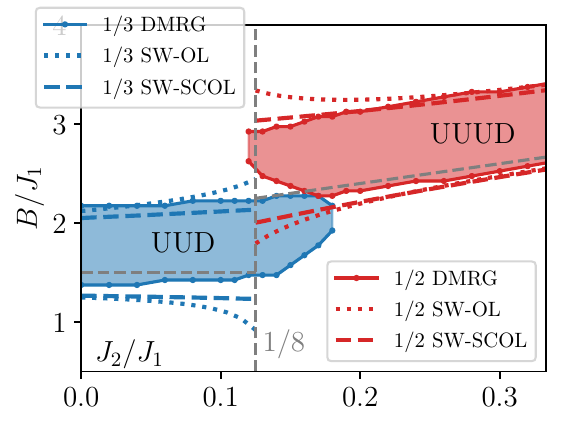}
    \llap{\parbox{0pt}{\vspace{-180pt}\hspace{-230pt}\footnotesize{(b)}}} 
    \includegraphics[width=0.48\linewidth]{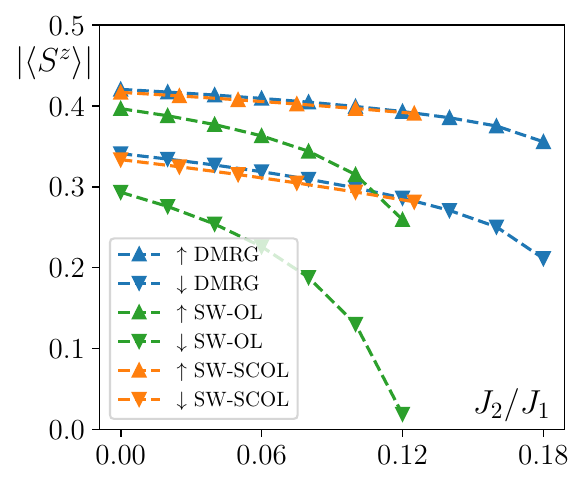}
    \llap{\parbox{0pt}{\vspace{-195pt}\hspace{-230pt}\footnotesize{(c)}}}
    \includegraphics[width=0.48\linewidth]{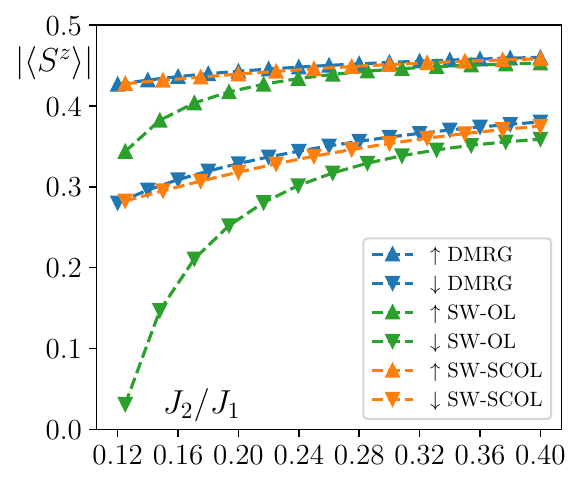}
    \llap{\parbox{0pt}{\vspace{-195pt}\hspace{-230pt}\footnotesize{(d)}}}
    \caption{(a) Magnetization as a function of magnetic field for different ratios of $J_2/J_1$ (offset for clarity) featuring the 1/3 and 1/2 plateaux. (b) Phase boundaries of the magnetization plateaux as obtained using DMRG (solid line) and spin waves (SW) one-loop (OL) and self-consistent one-loop (SCOL) analysis (dotted and dashed lines, respectively). Dashed gray line corresponds to $J_2/J_1=1/8$, indicating the phase boundaries of the plateaux in the classical limit. (c) Local magnetic moment in the 1/3 and (d) 1/2 magnetization plateaux. SW calculations are carried out at $1/3$ ($1/2$) of the saturation field. In DMRG total magnetization is fixed to the respective fraction.}
    \label{fig:Plateaux}
\end{figure}

\emph{Magnetization Plateaux. -}
A key feature of the phase diagram, 
predicted in~\cite{Ye2017}, and confirmed by our simulations is the appearance of the 1/2 magnetization plateau for large $J_2$.
In addition, we find that the well-known 1/3 plateau extends well beyond the first-order phase transition line expected at $J_2/J_1=1/8$. This results in an overlap of the two plateaux in the QSL range of $J_2$ values. This is demonstrated explicitly in the magnetization curves shown in Fig.~\ref{fig:Plateaux}(a).

In Fig.~\ref{fig:Plateaux}(b), we compare plateaux boundaries obtained using DMRG and the semiclassical spin-wave approach (see End Matter for details). 
The first approach consists of linear spin-wave theory (LSWT) supplemented by a one-loop (OL) correction. 
For collinear phases such as the UUD and UUUD states, cubic terms are not present and this correction is obtained by normal ordering the quartic interaction term, which renormalizes the quadratic spin-wave Hamiltonian. 
In the simple OL scheme, only the diagonal elements of the dynamical matrix are renormalized, whereas in the self-consistent OL (SCOL)  approach, all matrix elements are renormalized and the corresponding vacuum expectation values are determined self-consistently (see the End Matter and Sec.~III of the Supplemental Material (SM)~\cite{SM}).

To further characterize the two plateaux, we compute the local ordered moments on the two inequivalent sublattices using both iDMRG and spin-wave theory, as shown in Fig.~\ref{fig:Plateaux}(c,d). The one-loop spin-wave approximation already captures the qualitative behavior, including the suppression of the ordered moment as $J_2$ approaches the first-order transition. The self-consistent scheme, however, yields nearly-perfect quantitative agreement with the DMRG data. The reduction of the magnetic moment, $\Delta S^z_{\bf r} = 1/2 - |\langle S^z_{\bf r}\rangle|$, on the down sublattice in the UUD (UUUD) phase is exactly twice (three times) that of each up sublattice, ensuring that the net magnetization remains one-third (one-half) of the saturation value.

\begin{figure}
    \centering
    \includegraphics[width=0.43\linewidth]{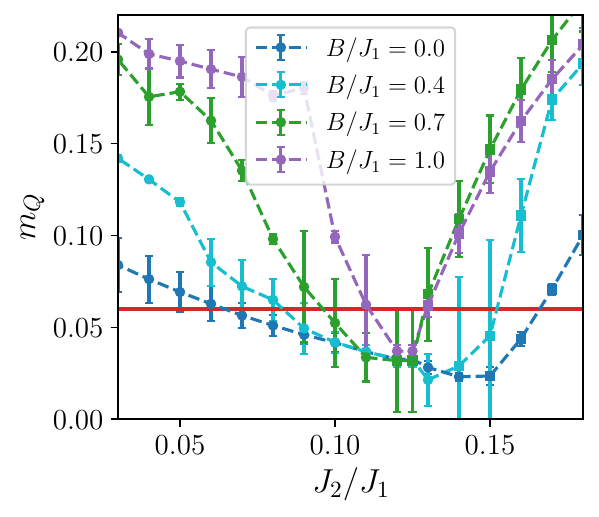}
    \llap{\parbox{0pt}{\vspace{-190pt}\hspace{-200pt}\footnotesize{(a)}}} 
    \includegraphics[trim={0 -0.8cm 0 0},clip, width=0.54\linewidth]{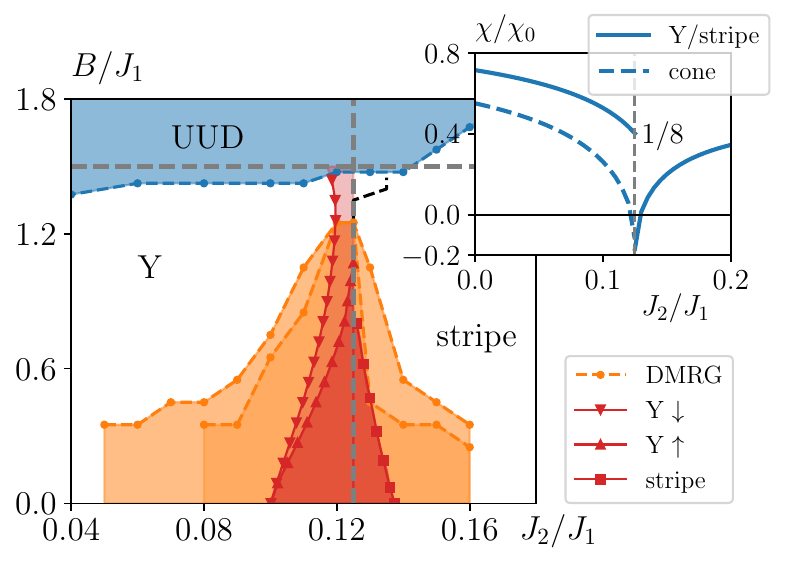}
    \llap{\parbox{0pt}{\vspace{-212pt}\hspace{-20pt}\footnotesize{(b)}}} 
    \caption{ (a) An estimate for the local ordered moment $m_{\bf Q}$  obtained using DMRG (see text for details) as a function of $J_2/J_1$ for different magnetic fields. Values obtained for ${\bf Q}={\bf K}$ (${\bf Q}={\bf M}$) are plotted as circles (squares). Error bars correspond to errors in extrapolation with $1/N_y$. The threshold value $m_{\rm th}=0.06$ used as an indicator of the quantum-disordered regime is plotted as a red horizontal line. (b) QSL region obtained using DMRG (orange filled area) and spin-wave theory (red filled area). The area bounded by the red dashed curves with up (down) triangle markers corresponds to the region in which the local ordered moment on the `tilted' (down) sublattice in the Y phase vanishes within linear spin-wave analysis. Similarly, the area bounded by the red dashed curve with square markers corresponds to the region in which the local ordered moment vanishes in the stripe phase. Inset shows the zero-field susceptibility of the Y and stripe states (solid line) as well as the cone state (dashed line) as a function of $J_2$ obtained within the spin-wave analysis. Here $\chi_0=1/9$ is the classical value for the susceptibility in the Y phase.}
    \label{fig:QSLboundaries}
\end{figure}

\emph{Vicinity of the QSL.—}
We now turn to the low-field region surrounding the quantum-disordered phase and outline the numerical procedure used to determine the QSL boundaries. The disordered state is characterized by a vanishing ordered moment, which we estimate via $m_{\mathbf{Q}}^2 = S(\mathbf{Q})/N$, where $\mathbf{Q}$ runs over the candidate ordering wave vectors. Computing $m_{\mathbf{Q}}$ on cylinders of circumference $N_y=4,6,8$ and extrapolating to $1/N_y\!\to\!0$, we obtain the thermodynamic ordered moment. Because $m_{\mathbf{Q}}$ is contaminated by short-range correlations at finite $N$, it remains nonzero even inside the QSL. At zero field we find that, throughout the established QSL window $0.06\!\lesssim\!J_2/J_1\!\lesssim\!0.16$, the maximal $m_{\mathbf{Q}}$ (taken between Y and stripe wave vectors) is suppressed below a threshold value $m_{\rm th}\!\sim\!0.06$. We use this threshold to identify the disordered regime at finite fields (Fig.~\ref{fig:QSLboundaries}(a)). Error bars reflect the extrapolation uncertainty, and the shaded regions in Fig.~\ref{fig:PhaseDiagram} and Fig.~\ref{fig:QSLboundaries}(b) indicate conservative and more permissive estimates of the QSL boundaries. Specifically, the lower (upper) critical field is taken as the largest (smallest) $B$ satisfying $m_{\mathbf{Q}}+\Delta m_{\mathbf{Q}}<m_{\rm th}$ ($m_{\mathbf{Q}}-\Delta m_{\mathbf{Q}}>m_{\rm th}$).

A complementary estimate follows from linear spin-wave theory. We compute the sublattice magnetizations $\langle S^z_{\mathbf{r}}\rangle$ for the Y and stripe phases, where quantum fluctuations reduce the classical value $S$ by the ground-state magnon occupation $\langle \hat{n}_{\mathbf{r}}\rangle = \Delta S^z_{\bf r}$.  
For $S=1/2$, long-range order persists as long as $\langle \hat{n}_{\mathbf{r}}\rangle \le 1/2$. Details of this calculation are given in the End Matter and Secs.~I and II of the SM~\cite{SM}. The resulting LRO boundaries (Fig.~\ref{fig:QSLboundaries}(b)) predict a zero-field disordered window $0.10\!\le\!J_2/J_1\!\le\!0.137$. They further show that fields of order $B\!\sim\!J_1$ are required to stabilize order near the first-order line at $J_2/J_1=1/8$. On the Y side, the down-sublattice remains disordered up to the $m=1/3$ plateau at $B\!=\!1.5J_1$, consistent with the constraint that at the plateau $\langle \hat{n}_{\downarrow}\rangle = 2\langle \hat{n}_{\uparrow}\rangle$. 
This criterion—requiring the ``most ordered’’ sublattice to satisfy $\langle \hat{n}\rangle < 1/2$—accounts for the spin-wave boundaries separating the Y and stripe phases from the QSL.

We next examine the uniform linear susceptibility $\chi = (\mathrm{d}m/\mathrm{d}B)_{B=0}$, which provides an additional diagnostic of magnetic order: the phase with the largest $\chi$ gains the most energy from an infinitesimal field and is therefore realized in the $B\!\to\!0$ limit. Using the three-sublattice spin-wave formalism of Refs.~\cite{Chubukov_1991,Chubukov1994}, we find that the coplanar Y state consistently outcompetes the noncoplanar cone (umbrella) state throughout its classical stability window $0 \le J_2/J_1 \le 1/8$. Quantum fluctuations reduce $\chi_{\rm Y}$ by roughly $40\%$ relative to its $J_2=0$ value, but it 
remains positive for all $J_2$ in this range (see the End Matter and Sec.~I of the SM~\cite{SM}). In contrast, the cone susceptibility becomes negative as $J_2/J_1 \to 1/8$, signaling an instability of the noncoplanar order [see inset of Fig.~\ref{fig:QSLboundaries}(b)]. On the stripe side, evaluating $\chi_{\rm stripe}$ in the rotated basis of Ref.~\cite{Mourigal2010} — which requires only a single boson species (see the End Matter and Sec.~II of the SM~\cite{SM}) — likewise shows that quantum fluctuations drive $\chi_{\rm stripe}$ negative near the classical transition at $J_2/J_1=1/8$, indicating the breakdown of the canted stripe state. 

These results are consistent with a quantum-disordered state
centered around the classical first-order line at $J_2/J_1 = 1/8$, and further indicate that this state remains remarkably robust under a magnetic field. To probe the nature of the magnetized QSL, which has been proposed to develop finite scalar chirality $\chi_\triangle = \mathbf{S}_i\!\cdot\!\left( \mathbf{S}_j \times \mathbf{S}_k \right)$~\cite{Ran2009,Chen2025qed3}, we computed scalar chirality correlations using iDMRG. On XC-8 cylinders, the correlations decay rapidly throughout the QSL regime (see End Matter). On YC-6 cylinders, however, we find a competing noncoplanar cone state with long-range chirality correlations in a region neighboring and partially overlapping the QSL phase. We attribute this cone state and its finite chirality to finite-size effects (see End Matter).

\emph{High magnetization regime.—}
We now turn to the vicinity of saturation. 
Ref.~\cite{Ye2017} suggested that the stripe state at high fields is suppressed near $J_2/J_1 = 1/3$, where magnons condense simultaneously at all three $M$ points, yielding the $\bar{V}$ phase instead of the single-$M$ condensate characteristic of the stripe state that breaks $C_3$ symmetry. Carrying out a magnon condensation analysis which is exact for $S=1/2$ (see Sec.IV~\cite{SM} for details), we fully confirm this prediction. Our numerical results further corroborate this finding: the stripe phase occupies a narrow field window near saturation, but is pushed out by the $\bar{V}$ phase near $J_2/J_1\!\approx\!1/3$.

Away from $J_2/J_1 = 1/3$, our numerics reveal a stripe supersolid with oscillations in both longitudinal and transverse spin components. However, the stability range of this phase depends strongly on the iDMRG boundary conditions, suggesting that it is likely an artifact of the cylindrical geometry, which explicitly breaks the lattice $C_3$ symmetry. Nevertheless, such a phase could naturally arise in systems with intrinsic lattice anisotropy, as discussed in the End Matter.

Ref.~\cite{Ye2017} also predicted a quantum-disordered state near saturation around the first-order line $J_2/J_1 = 1/8$. While our numerics indicate that the system may remain disordered extremely close to this point (see End Matter), we find no evidence for an extended disordered regime. Distinguishing this behavior from very weak magnetic order near the transition remains challenging.

\emph{Discussion. -} 
Our quantum phase diagram identifies the coexistence of the $m=1/3$ and $m=1/2$ magnetization plateaux as a robust signature of proximity to the quantum spin-liquid regime. More importantly, our results show that the strong quantum fluctuations associated with this regime remain remarkably resilient against magnetic fields up to $B \gtrsim J_1$, despite the emergence of field-induced ordered states at higher magnetization. This robustness is particularly significant given the ongoing debate regarding the precise nature of the zero-field state~\cite{jiang2026,kovalska2026}. Regardless of whether the zero-field phase is ultimately identified as a Dirac QSL or another competing nonmagnetic state, our results demonstrate that its proximity leaves clear, experimentally accessible fingerprints in the field dependence of the magnetization process. In particular, the coexistence of the $m=1/3$ and $m=1/2$ plateaux provides a practical diagnostic for identifying regions of the phase diagram dominated by strong quantum effects rather than semiclassical ordering tendencies.

Existing measurements on powder samples of NaYbO$_2$~\cite{Wilson2019,Tsirlin2019} and KYbO$_2$~\cite{Tsirlin2023} provide incipient, although not yet conclusive, indications of the plateau phenomenology predicted in our work. In NaYbO$_2$, neutron scattering has identified field-induced UUD order associated with the 1/3 plateau \cite{Wilson2019}, while a separate magnetization study revealed a narrow plateau-like feature near one-half of the saturation magnetization \cite{Tsirlin2019}. A 1/2 plateau has also been reported in KYbO$_2$ \cite{Tsirlin2023}. We call for further {\em single-crystal} magnetization and field-resolved inelastic neutron scattering studies of these and other nearly isotropic spin-1/2 triangular materials \cite{shimizu2026design}.

The excellent agreement between DMRG results and the self-consistent one-loop calculation for the stability range of each plateau further demonstrates that the same semiclassical framework can be reliably employed to extract microscopic Hamiltonian parameters from fits to inelastic neutron scattering data~\cite{Xie2023,KamiyaY2018a,Zhang2026}.

Our DMRG study provides approximate phase boundaries for the QSL in finite field. However, determining the precise nature of this state requires probing its low-energy excitations using flux threading analysis~\cite{He2017Kagome,He2019}, which we leave for future investigation.

\emph{Note added.-}  After completion of this work, we learned of an independent complementary study~\cite{Bader2026-prb} that investigates the same problem and reports conclusions mostly consistent with ours.

\emph{Acknowledgments. -} We thank Federico Becca, Sasha Chernyshev, Yasir Iqbal, Johannes Knolle, Frank Pollmann, Mengxing Ye, and Stephen Wilson for fruitful discussions.
A.K. and O.A.S. are supported by Grant No. 2024200 from the United States-Israel Binational Science Foundation (BSF).
A.K.\ acknowledges funding by the Israeli Council for Higher Education support program and by the Israel Science Foundation (Grant No.\ 2443/22). 
Work at the University of Tennessee was
supported by National Science Foundation Materials Research
Science and Engineering Center program through the UT
Knoxville Center for Advanced Materials and Manufacturing
(DMR-2309083) and by the Lincoln Chair of Excellence in
Physics. 


\let\oldaddcontentsline\addcontentsline
\renewcommand{\addcontentsline}[3]{}
\bibliography{J1-J2-refs}

@article{Anderson1973,
title = {Resonating valence bonds: A new kind of insulator?},
journal = {Materials Research Bulletin},
volume = {8},
number = {2},
pages = {153-160},
year = {1973},
issn = {0025-5408},
doi = {https://doi.org/10.1016/0025-5408(73)90167-0},
url = {https://www.sciencedirect.com/science/article/pii/0025540873901670},
author = {P.W. Anderson}
}

@article{Zhang2026,
  title={Nonperturbative semiclassical spin dynamics for ordered quantum magnets},
  author={Zhang, Hao and Huang, Tianyue and Scheie, Allen O and Zhu, Mengze and Xie, Tao and Murai, Naoki and Ohira-Kawamura, Seiko and Zheludev, Andrey and L{\"a}uchli, Andreas M and Batista, Cristian D},
  journal={npj Quantum Materials},
  year={2026},
  publisher={Nature Publishing Group UK London},
  doi={doi.org/10.1038/s41535-026-00873-9}
}

@article{Starykh2014,
  title = {Phases of a Triangular-Lattice Antiferromagnet Near Saturation},
  author = {Starykh, Oleg A. and Jin, Wen and Chubukov, Andrey V.},
  journal = {Phys. Rev. Lett.},
  volume = {113},
  issue = {8},
  pages = {087204},
  numpages = {5},
  year = {2014},
  month = {Aug},
  publisher = {American Physical Society},
  doi = {10.1103/PhysRevLett.113.087204},
  url = {https://link.aps.org/doi/10.1103/PhysRevLett.113.087204}
}

@Article{Xie2023,
author={Xie, Tao
and Eberharter, A. A.
and Xing, Jie
and Nishimoto, S.
and Brando, M.
and Khanenko, P.
and Sichelschmidt, J.
and Turrini, A. A.
and Mazzone, D. G.
and Naumov, P. G.
and Sanjeewa, L. D.
and Harrison, N.
and Sefat, Athena S.
and Normand, B.
and L{\"a}uchli, A. M.
and Podlesnyak, A.
and Nikitin, S. E.},
title={{Complete field-induced spectral response of the spin-1/2 triangular-lattice antiferromagnet CsYbSe$_2$}},
journal={npj Quantum Materials},
year={2023},
month={Sep},
day={23},
volume={8},
number={1},
pages={48},
issn={2397-4648},
doi={10.1038/s41535-023-00580-9},
url={https://doi.org/10.1038/s41535-023-00580-9}
}

@article{Haravifard2024,
  title = {{Evidence of Dirac Quantum Spin Liquid in ${\mathrm{YbZn}}_{2}{\mathrm{GaO}}_{5}$}},
  author = {Bag, Rabindranath and Xu, Sijie and Sherman, Nicholas E. and Yadav, Lalit and Kolesnikov, Alexander I. and Podlesnyak, Andrey A. and Choi, Eun Sang and da Silva, Ivan and Moore, Joel E. and Haravifard, Sara},
  journal = {Phys. Rev. Lett.},
  volume = {133},
  issue = {26},
  pages = {266703},
  numpages = {10},
  year = {2024},
  month = {Dec},
  publisher = {American Physical Society},
  doi = {10.1103/PhysRevLett.133.266703},
  url = {https://link.aps.org/doi/10.1103/PhysRevLett.133.266703}
}

@article{Scheie2024,
  title = {{Nonlinear magnons and exchange Hamiltonians of the delafossite proximate quantum spin liquid candidates ${\text{KYbSe}}_{2}$ and ${\text{NaYbSe}}_{2}$}},
  author = {Scheie, A. O. and Kamiya, Y. and Zhang, Hao and Lee, Sangyun and Woods, A. J. and Ajeesh, M. O. and Gonzalez, M. G. and Bernu, B. and Villanova, J. W. and Xing, J. and Huang, Q. and Zhang, Qingming and Ma, Jie and Choi, Eun Sang and Pajerowski, D. M. and Zhou, Haidong and Sefat, A. S. and Okamoto, S. and Berlijn, T. and Messio, L. and Movshovich, R. and Batista, C. D. and Tennant, D. A.},
  journal = {Phys. Rev. B},
  volume = {109},
  issue = {1},
  pages = {014425},
  numpages = {12},
  year = {2024},
  month = {Jan},
  publisher = {American Physical Society},
  doi = {10.1103/PhysRevB.109.014425},
  url = {https://link.aps.org/doi/10.1103/PhysRevB.109.014425}
}

@article{Imada2014,
author = {Kaneko ,Ryui and Morita ,Satoshi and Imada ,Masatoshi},
title = {{Gapless Spin-Liquid Phase in an Extended Spin 1/2 Triangular Heisenberg Model}},
journal = {Journal of the Physical Society of Japan},
volume = {83},
number = {9},
pages = {093707},
year = {2014},
doi = {10.7566/JPSJ.83.093707}
}

@article{Bishop2015,
  title = {{Quasiclassical magnetic order and its loss in a spin-$\frac{1}{2}$ Heisenberg antiferromagnet on a triangular lattice with competing bonds}},
  author = {Li, P. H. Y. and Bishop, R. F. and Campbell, C. E.},
  journal = {Phys. Rev. B},
  volume = {91},
  issue = {1},
  pages = {014426},
  numpages = {11},
  year = {2015},
  month = {Jan},
  publisher = {American Physical Society},
  doi = {10.1103/PhysRevB.91.014426},
  url = {https://link.aps.org/doi/10.1103/PhysRevB.91.014426}
}

@article{White2015,
  title = {{Spin liquid phase of the $S=\frac{1}{2}\phantom{\rule{4.pt}{0ex}}{J}_{1}\ensuremath{-}{J}_{2}$ Heisenberg model on the triangular lattice}},
  author = {Zhu, Zhenyue and White, Steven R.},
  journal = {Phys. Rev. B},
  volume = {92},
  issue = {4},
  pages = {041105},
  numpages = {4},
  year = {2015},
  month = {Jul},
  publisher = {American Physical Society},
  doi = {10.1103/PhysRevB.92.041105},
  url = {https://link.aps.org/doi/10.1103/PhysRevB.92.041105}
}

@article{Hu2015,
  title = {{Competing spin-liquid states in the spin-$\frac{1}{2}$ Heisenberg model on the triangular lattice}},
  author = {Hu, Wen-Jun and Gong, Shou-Shu and Zhu, Wei and Sheng, D. N.},
  journal = {Phys. Rev. B},
  volume = {92},
  issue = {14},
  pages = {140403},
  numpages = {6},
  year = {2015},
  month = {Oct},
  publisher = {American Physical Society},
  doi = {10.1103/PhysRevB.92.140403},
  url = {https://link.aps.org/doi/10.1103/PhysRevB.92.140403}
}

@article{Iqbal2016,
  title = {{Spin liquid nature in the Heisenberg ${J}_{1}\ensuremath{-}{J}_{2}$ triangular antiferromagnet}},
  author = {Iqbal, Yasir and Hu, Wen-Jun and Thomale, Ronny and Poilblanc, Didier and Becca, Federico},
  journal = {Phys. Rev. B},
  volume = {93},
  issue = {14},
  pages = {144411},
  numpages = {14},
  year = {2016},
  month = {Apr},
  publisher = {American Physical Society},
  doi = {10.1103/PhysRevB.93.144411},
  url = {https://link.aps.org/doi/10.1103/PhysRevB.93.144411}
}

@article{Gong2017,
  title = {Global phase diagram and quantum spin liquids in a spin-$\frac{1}{2}$ triangular antiferromagnet},
  author = {Gong, Shou-Shu and Zhu, W. and Zhu, J.-X. and Sheng, D. N. and Yang, Kun},
  journal = {Phys. Rev. B},
  volume = {96},
  issue = {7},
  pages = {075116},
  numpages = {10},
  year = {2017},
  month = {Aug},
  publisher = {American Physical Society},
  doi = {10.1103/PhysRevB.96.075116},
  url = {https://link.aps.org/doi/10.1103/PhysRevB.96.075116}
}

@article{Saadamand2017,
  title = {{Detection and characterization of symmetry-broken long-range orders in the spin-$\frac{1}{2}$ triangular Heisenberg model}},
  author = {Saadatmand, S. N. and McCulloch, I. P.},
  journal = {Phys. Rev. B},
  volume = {96},
  issue = {7},
  pages = {075117},
  numpages = {21},
  year = {2017},
  month = {Aug},
  publisher = {American Physical Society},
  doi = {10.1103/PhysRevB.96.075117},
  url = {https://link.aps.org/doi/10.1103/PhysRevB.96.075117}
}

@article{He2019,
  title = {{Dirac Spin Liquid on the Spin-$1/2$ Triangular Heisenberg Antiferromagnet}},
  author = {Hu, Shijie and Zhu, W. and Eggert, Sebastian and He, Yin-Chen},
  journal = {Phys. Rev. Lett.},
  volume = {123},
  issue = {20},
  pages = {207203},
  numpages = {6},
  year = {2019},
  month = {Nov},
  publisher = {American Physical Society},
  doi = {10.1103/PhysRevLett.123.207203},
  url = {https://link.aps.org/doi/10.1103/PhysRevLett.123.207203}
}

@article{He2017Kagome,
  title = {{Signatures of Dirac Cones in a DMRG Study of the Kagome Heisenberg Model}},
  author = {He, Yin-Chen and Zaletel, Michael P. and Oshikawa, Masaki and Pollmann, Frank},
  journal = {Phys. Rev. X},
  volume = {7},
  issue = {3},
  pages = {031020},
  numpages = {16},
  year = {2017},
  month = {Jul},
  publisher = {American Physical Society},
  doi = {10.1103/PhysRevX.7.031020},
  url = {https://link.aps.org/doi/10.1103/PhysRevX.7.031020}
}

@article{Gallegos2025,
  title = {Phase Diagram of the Easy-Axis Triangular-Lattice ${J}_{1}\text{\ensuremath{-}}{J}_{2}$ Model},
  author = {Gallegos, Cesar A. and Jiang, Shengtao and White, Steven R. and Chernyshev, A. L.},
  journal = {Phys. Rev. Lett.},
  volume = {134},
  issue = {19},
  pages = {196702},
  numpages = {9},
  year = {2025},
  month = {May},
  publisher = {American Physical Society},
  doi = {10.1103/PhysRevLett.134.196702},
  url = {https://link.aps.org/doi/10.1103/PhysRevLett.134.196702}
}

@article{Oitmaa2020,
  title = {{Magnetic phases in the ${J}_{1}\text{\ensuremath{-}}{J}_{2}$ Heisenberg antiferromagnet on the triangular lattice}},
  author = {Oitmaa, J.},
  journal = {Phys. Rev. B},
  volume = {101},
  issue = {21},
  pages = {214422},
  numpages = {5},
  year = {2020},
  month = {Jun},
  publisher = {American Physical Society},
  doi = {10.1103/PhysRevB.101.214422},
  url = {https://link.aps.org/doi/10.1103/PhysRevB.101.214422}
}

@article{Wietek2024,
  title = {Quantum Electrodynamics in $2+1$ Dimensions as the Organizing Principle of a Triangular Lattice Antiferromagnet},
  author = {Wietek, Alexander and Capponi, Sylvain and L\"auchli, Andreas M.},
  journal = {Phys. Rev. X},
  volume = {14},
  issue = {2},
  pages = {021010},
  numpages = {16},
  year = {2024},
  month = {Apr},
  publisher = {American Physical Society},
  doi = {10.1103/PhysRevX.14.021010},
  url = {https://link.aps.org/doi/10.1103/PhysRevX.14.021010}
}

@Article{Song2019,
author={Song, Xue-Yang
and Wang, Chong
and Vishwanath, Ashvin
and He, Yin-Chen},
title={Unifying description of competing orders in two-dimensional quantum magnets},
journal={Nature Communications},
year={2019},
month={Sep},
day={18},
volume={10},
number={1},
pages={4254},
issn={2041-1723},
doi={10.1038/s41467-019-11727-3},
url={https://doi.org/10.1038/s41467-019-11727-3}
}

@article{Hermele2005,
  title = {Algebraic spin liquid as the mother of many competing orders},
  author = {Hermele, Michael and Senthil, T. and Fisher, Matthew P. A.},
  journal = {Phys. Rev. B},
  volume = {72},
  issue = {10},
  pages = {104404},
  numpages = {16},
  year = {2005},
  month = {Sep},
  publisher = {American Physical Society},
  doi = {10.1103/PhysRevB.72.104404},
  url = {https://link.aps.org/doi/10.1103/PhysRevB.72.104404}
}

@article{Jolicoeur1990,
  title = {{Ground-state properties of the S=1/2 Heisenberg antiferromagnet on a triangular lattice}},
  author = {Jolicoeur, Th. and Dagotto, E. and Gagliano, E. and Bacci, S.},
  journal = {Phys. Rev. B},
  volume = {42},
  issue = {7},
  pages = {4800--4803},
  numpages = {0},
  year = {1990},
  month = {Sep},
  publisher = {American Physical Society},
  doi = {10.1103/PhysRevB.42.4800},
  url = {https://link.aps.org/doi/10.1103/PhysRevB.42.4800}
}

@article{Chubukov1992,
  title = {{Order-from-disorder phenomena in Heisenberg antiferromagnets on a triangular lattice}},
  author = {Chubukov, Andrey V. and Jolicoeur, Th.},
  journal = {Phys. Rev. B},
  volume = {46},
  issue = {17},
  pages = {11137--11140},
  numpages = {0},
  year = {1992},
  month = {Nov},
  publisher = {American Physical Society},
  doi = {10.1103/PhysRevB.46.11137},
  url = {https://link.aps.org/doi/10.1103/PhysRevB.46.11137}
}

@article{Korshynov1993,
  title = {{Chiral phase of the Heisenberg antiferromagnet with a triangular lattice}},
  author = {Korshunov, S. E.},
  journal = {Phys. Rev. B},
  volume = {47},
  issue = {10},
  pages = {6165--6168},
  numpages = {0},
  year = {1993},
  month = {Mar},
  publisher = {American Physical Society},
  doi = {10.1103/PhysRevB.47.6165},
  url = {https://link.aps.org/doi/10.1103/PhysRevB.47.6165}
}

@article{Chubukov_1991,
doi = {10.1088/0953-8984/3/1/005},
url = {https://dx.doi.org/10.1088/0953-8984/3/1/005},
year = {1991},
month = {jan},
publisher = {},
volume = {3},
number = {1},
pages = {69},
author = {A V Chubukov and D I Golosov},
title = {Quantum theory of an antiferromagnet on a triangular lattice in a magnetic field},
journal = {Journal of Physics: Condensed Matter}
}

@article{Starykh_2015,
doi = {10.1088/0034-4885/78/5/052502},
url = {https://dx.doi.org/10.1088/0034-4885/78/5/052502},
year = {2015},
month = {apr},
publisher = {IOP Publishing},
volume = {78},
number = {5},
pages = {052502},
author = {Starykh, Oleg A},
title = {Unusual ordered phases of highly frustrated magnets: a review},
journal = {Reports on Progress in Physics}
}

@article{Ye2017,
  title = {{Quantum phase transitions in the Heisenberg ${J}_{1}\ensuremath{-}{J}_{2}$ triangular antiferromagnet in a magnetic field}},
  author = {Ye, Mengxing and Chubukov, Andrey V.},
  journal = {Phys. Rev. B},
  volume = {95},
  issue = {1},
  pages = {014425},
  numpages = {27},
  year = {2017},
  month = {Jan},
  publisher = {American Physical Society},
  doi = {10.1103/PhysRevB.95.014425},
  url = {https://link.aps.org/doi/10.1103/PhysRevB.95.014425}
}

@article{Ye2017b,
  title = {{Half-magnetization plateau in a Heisenberg antiferromagnet on a triangular lattice}},
  author = {Ye, Mengxing and Chubukov, Andrey V.},
  journal = {Phys. Rev. B},
  volume = {96},
  issue = {14},
  pages = {140406},
  numpages = {5},
  year = {2017},
  month = {Oct},
  publisher = {American Physical Society},
  doi = {10.1103/PhysRevB.96.140406},
  url = {https://link.aps.org/doi/10.1103/PhysRevB.96.140406}
}

@article{Tsirlin_2020,
doi = {10.1088/1361-648X/ab724e},
url = {https://dx.doi.org/10.1088/1361-648X/ab724e},
year = {2020},
month = {mar},
publisher = {IOP Publishing},
volume = {32},
number = {22},
pages = {224004},
author = {Li, Yuesheng and Gegenwart, Philipp and Tsirlin, Alexander A},
title = {Spin liquids in geometrically perfect triangular antiferromagnets},
journal = {Journal of Physics: Condensed Matter}
}

@article{Chubukov1994,
doi = {10.1088/0953-8984/6/42/019},
url = {https://dx.doi.org/10.1088/0953-8984/6/42/019},
year = {1994},
month = {oct},
publisher = {},
volume = {6},
number = {42},
pages = {8891},
author = {A V Chubukov and S Sachdev and T Senthil},
title = {{Large-S expansion for quantum antiferromagnets on a triangular lattice}},
journal = {Journal of Physics: Condensed Matter}
}

@article{Mourigal2010,
  title = {Field-induced decay dynamics in square-lattice antiferromagnets},
  author = {Mourigal, M. and Zhitomirsky, M. E. and Chernyshev, A. L.},
  journal = {Phys. Rev. B},
  volume = {82},
  issue = {14},
  pages = {144402},
  numpages = {12},
  year = {2010},
  month = {Oct},
  publisher = {American Physical Society},
  doi = {10.1103/PhysRevB.82.144402},
  url = {https://link.aps.org/doi/10.1103/PhysRevB.82.144402}
}

@article{Chernyshev2022,
  title = {{Roller Coaster in a Flatland: Magnetoresistivity in Eu-Intercalated Graphite}},
  author = {Chernyshev, A. L. and Starykh, O. A.},
  journal = {Phys. Rev. X},
  volume = {12},
  issue = {2},
  pages = {021010},
  numpages = {37},
  year = {2022},
  month = {Apr},
  publisher = {American Physical Society},
  doi = {10.1103/PhysRevX.12.021010},
  url = {https://link.aps.org/doi/10.1103/PhysRevX.12.021010}
}

@article{Krivoruchko1981,
author = {Krivoruchko, V. N. and Yablonskii, D. A.},
title = {{On the Problem of Calculating the Bogolyubov U-VTransformation Coefficients}},
journal = {physica status solidi (b)},
volume = {103},
number = {1},
pages = {K41-K45},
doi = {https://doi.org/10.1002/pssb.2221030159},
url = {https://onlinelibrary.wiley.com/doi/abs/10.1002/pssb.2221030159},
year = {1981}
}

@article{Maksimov2019,
  title = {Anisotropic-Exchange Magnets on a Triangular Lattice: Spin Waves, Accidental Degeneracies, and Dual Spin Liquids},
  author = {Maksimov, P. A. and Zhu, Zhenyue and White, Steven R. and Chernyshev, A. L.},
  journal = {Phys. Rev. X},
  volume = {9},
  issue = {2},
  pages = {021017},
  numpages = {27},
  year = {2019},
  month = {Apr},
  publisher = {American Physical Society},
  doi = {10.1103/PhysRevX.9.021017},
  url = {https://link.aps.org/doi/10.1103/PhysRevX.9.021017}
}

@article{Veillette2005,
  title = {{Ground states of a frustrated spin-$\frac{1}{2}$ antiferromagnet: ${\mathrm{Cs}}_{2}\mathrm{Cu}{\mathrm{Cl}}_{4}$ in a magnetic field}},
  author = {Veillette, M. Y. and Chalker, J. T. and Coldea, R.},
  journal = {Phys. Rev. B},
  volume = {71},
  issue = {21},
  pages = {214426},
  numpages = {13},
  year = {2005},
  month = {Jun},
  publisher = {American Physical Society},
  doi = {10.1103/PhysRevB.71.214426},
  url = {https://link.aps.org/doi/10.1103/PhysRevB.71.214426}
}

@Article{tenpy2024,
    title={{Tensor network Python (TeNPy) version 1}},
    author={Johannes Hauschild and Jakob Unfried and Sajant Anand and Bartholomew Andrews and Marcus Bintz and Umberto Borla and Stefan Divic and Markus Drescher and Jan Geiger and Martin Hefel and Kévin Hémery and Wilhelm Kadow and Jack Kemp and Nico Kirchner and Vincent S. Liu and Gunnar Möller and Daniel Parker and Michael Rader and Anton Romen and Samuel Scalet and Leon Schoonderwoerd and Maximilian Schulz and Tomohiro Soejima and Philipp Thoma and Yantao Wu and Philip Zechmann and Ludwig Zweng and Roger S. K. Mong and Michael P. Zaletel and Frank Pollmann},
    journal={SciPost Phys. Codebases},
    volume = {41},
    year={2024},
    publisher={SciPost},
    doi={10.21468/SciPostPhysCodeb.41},
    url={https://scipost.org/10.21468/SciPostPhysCodeb.41},
}

@article{Ran2009,
  title = {{Spontaneous Spin Ordering of a Dirac Spin Liquid in a Magnetic Field}},
  author = {Ran, Ying and Ko, Wing-Ho and Lee, Patrick A. and Wen, Xiao-Gang},
  journal = {Phys. Rev. Lett.},
  volume = {102},
  issue = {4},
  pages = {047205},
  numpages = {4},
  year = {2009},
  month = {Jan},
  publisher = {American Physical Society},
  doi = {10.1103/PhysRevLett.102.047205},
  url = {https://link.aps.org/doi/10.1103/PhysRevLett.102.047205}
}

@article{Chen2025qed3,
  title = {{Emergent gauge flux in QED${}_3$ with flavor chemical potential: application to magnetized {U}(1) Dirac spin liquids}},
  author = {Chen, Chuang and Seifert, Urban F. P. and Feng, Kexin and Starykh, Oleg A. and Balents, Leon and Meng, Zi Yang},
  journal      = {arXiv preprint},
  year         = {2025},
  doi          = {10.48550/arXiv.2508.08528},
}

@article{AliceaJ2009,
  author = {Alicea, Jason and Chubukov, Andrey V. and Starykh, Oleg A.},
  year = {2009},
  month = mar,
  journal = {Physical Review Letters},
  volume = {102},
  number = {13},
  pages = {137201},
  publisher = {American Physical Society},
  doi = {10.1103/PhysRevLett.102.137201},
  url = {https://link.aps.org/doi/10.1103/PhysRevLett.102.137201},
  urldate = {2025-05-22},
  title = {Quantum {{Stabilization}} of the {$1/3$}-{{Magnetization Plateau}} in {${\mathrm{Cs}}_{2}{\mathrm{CuBr}}_{4}$}}
}

@article{KamiyaY2018a,
  title = {The Nature of Spin Excitations in the One-Third Magnetization Plateau Phase of {{Ba$_3$CoSb$_2$O$_9$}}},
  author = {Kamiya, Y. and Ge, L. and Hong, Tao and Qiu, Y. and {Quintero-Castro}, D. L. and Lu, Z. and Cao, H. B. and Matsuda, M. and Choi, E. S. and Batista, C. D. and Mourigal, M. and Zhou, H. D. and Ma, J.},
  year = {2018},
  month = jul,
  journal = {Nature Communications},
  volume = {9},
  number = {1},
  pages = {2666},
  publisher = {Nature Publishing Group},
  issn = {2041-1723},
  doi = {10.1038/s41467-018-04914-1},
  url = {https://www.nature.com/articles/s41467-018-04914-1},
  urldate = {2025-02-12},
  copyright = {2018 The Author(s)},
  langid = {english},
}

@article{Sasank2025,
  title = {{Monopole excitations in the $U(1)$ Dirac spin liquid on the triangular lattice}},
  author = {Budaraju, Sasank and Parola, Alberto and Iqbal, Yasir and Becca, Federico and Poilblanc, Didier},
  journal = {Phys. Rev. B},
  volume = {111},
  issue = {12},
  pages = {125150},
  numpages = {10},
  year = {2025},
  month = {Mar},
  publisher = {American Physical Society},
  doi = {10.1103/PhysRevB.111.125150},
  url = {https://link.aps.org/doi/10.1103/PhysRevB.111.125150}
}

@misc{Zhang2026TriangularPlateaux,
  author = {Zhang, Hao},
  title = {{GitHub} repository for semiclassical spin-wave analysis of triangular-lattice magnetization plateaux},
  year = {2026},
  url = {https://github.com/Hao-Phys/TriangularPlateaux.jl},
}

@misc{Sunny_software,
  title = {Sunny.Jl {{GitHub Repository}}},
  author = {Dahlbom, David and Zhang, Hao and Miles, Cole and Quinn, Sam and Niraula, Alin and Thipe, Bhushan and Wilson, Matthew and Matin, Sakib and Mankad, Het and Hahn, Steven and Pajerowski, Daniel and Johnston, Steve and Wang, Zhentao and Lane, Harry and Li, Ying Wai and Bai, Xiaojian and Mourigal, Martin and Batista, Cristian D. and Barros, Kipton},
  year = {2025},
  month = may,
  url = {https://github.com/SunnySuite/Sunny.jl},
  urldate = {2025-05-24}
}

@article{Sunny_paper, 
doi = {10.21105/joss.08138}, 
url = {https://doi.org/10.21105/joss.08138}, 
year = {2025}, 
publisher = {The Open Journal}, 
volume = {10}, 
number = {116}, 
pages = {8138}, 
author = {Dahlbom, David and Zhang, Hao and Miles, Cole and Quinn, Sam and Niraula, Alin and Thipe, Bhushan and Wilson, Matthew and Matin, Sakib and Mankad, Het and Hahn, Steven and Pajerowski, Daniel and Johnston, Steve and Wang, Zhentao and Lane, Harry and Li, Ying Wai and Bai, Xiaojian and Mourigal, Martin and Batista, Cristian D. and Barros, Kipton}, 
title = {{Sunny.jl: A Julia Package for Spin Dynamics}}, journal = {Journal of Open Source Software} }

@article{ChernyshevAL2009,
  title = {{Spin Waves in a Triangular Lattice Antiferromagnet: Decays, Spectrum Renormalization, and Singularities}},
  shorttitle = {Spin Waves in a Triangular Lattice Antiferromagnet},
  author = {Chernyshev, A. L. and Zhitomirsky, M. E.},
  year = {2009},
  month = apr,
  journal = {Physical Review B},
  volume = {79},
  number = {14},
  pages = {144416},
  publisher = {American Physical Society},
  doi = {10.1103/PhysRevB.79.144416},
  url = {https://link.aps.org/doi/10.1103/PhysRevB.79.144416},
  urldate = {2025-05-22}
}

@article{Nikuni1995,
author = {Nikuni ,Tetsuro and Shiba ,Hiroyuki},
title = {{Hexagonal Antiferromagnets in Strong Magnetic Field:   Mapping onto Bose Condensation of Low-Density Bose Gas}},
journal = {Journal of the Physical Society of Japan},
volume = {64},
number = {9},
pages = {3471-3483},
year = {1995},
doi = {10.1143/JPSJ.64.3471}
}

@article{Batista2014,
  title = {{Bose-Einstein condensation in quantum magnets}},
  author = {Zapf, Vivien and Jaime, Marcelo and Batista, C. D.},
  journal = {Rev. Mod. Phys.},
  volume = {86},
  issue = {2},
  pages = {563--614},
  numpages = {52},
  year = {2014},
  month = {May},
  publisher = {American Physical Society},
  doi = {10.1103/RevModPhys.86.563},
  url = {https://link.aps.org/doi/10.1103/RevModPhys.86.563}
}

@article{jiang2026,
  title = {{Competing States in the $S=1/2$ Triangular-Lattice ${J}_{1}\text{\ensuremath{-}}{J}_{2}$ Heisenberg Model: A Dynamical Density-Matrix Renormalization Group Study}},
  author = {Jiang, Shengtao and White, Steven R. and Kivelson, Steven A. and Jiang, Hong-Chen},
  journal = {Phys. Rev. Lett.},
  volume = {137},
  issue = {5},
  pages = {056703},
  numpages = {10},
  year = {2026},
  month = {Jul},
  publisher = {American Physical Society},
  doi = {10.1103/zmnz-tkq2},
  url = {https://link.aps.org/doi/10.1103/zmnz-tkq2}
}

@article{kovalska2026,
  title = "{{Revisiting the $J_1$-$J_2$ Heisenberg Model on a Triangular Lattice: Quasi-Degenerate Ground States and Phase Competition}}",
  author = {{Kovalska}, Oleksandra and {Pag{\`e}s Fontanella}, Ester and {Schneider}, Benedikt and {Tu}, Hong-Hao and {von Delft}, Jan},
  journal      = {arXiv preprint},
  year         = {2026},
  doi          = {10.48550/arXiv.2603.08650},
}

@misc{SM,
  note = {See Supplemental Material for details on the spin-wave calculations, including one-loop and self-consistent one-loop corrections, as well as on a high-field analysis of magnon condensation for $J_2/J_1>1/8$, which includes Refs.~\cite{Zhang2026TriangularPlateaux,Sunny_software,Sunny_paper,ChernyshevAL2009,Nikuni1995,Batista2014}.},
}

@article{Bader2026-prb,
  title = {{Triangular ${J}_{1}\text{\ensuremath{-}}{J}_{2}$ Heisenberg antiferromagnet in a magnetic field}},
  author = {Bader, T. and Feng, S. and Budaraju, S. and Becca, F. and Knolle, J. and Pollmann, F.},
  journal = {Phys. Rev. B},
  volume = {113},
  issue = {24},
  pages = {L241114},
  numpages = {7},
  year = {2026},
  month = {Jun},
  publisher = {American Physical Society},
  doi = {10.1103/rwhz-jf5b},
  url = {https://link.aps.org/doi/10.1103/rwhz-jf5b}
}

@article{shimizu2026design,
  title={{Design Principles for Quasi-Isotropic Exchange in Rare-Earth Quantum Magnets}},
  author={Shimizu, Kotaro and Ghioldi, Esteban Agustin and Ronning, Filip and Batista, Cristian D},
  journal={arXiv preprint},
  doi={10.48550/arXiv.2606.31558}, 
  year={2026}
}

@article{Wilson2019,
  title={{Field-tunable quantum disordered ground state in the triangular-lattice antiferromagnet NaYbO$_2$}},
  author  = {Bordelon, Mitchell M. and Kenney, Eric and Liu, Chunxiao and Hogan, Tom and Posthuma, Lorenzo and Kavand, Marzieh and Lyu, Yuanqi and Sherwin, Mark and Butch, N. P. and Brown, Craig and Graf, M. J. and Balents, Leon and Wilson, Stephen D.},
  journal={Nature physics},
  volume={15},
  number={10},
  pages={1058--1064},
  year={2019},
  publisher={Nature Publishing Group UK London},
  doi={10.1038/s41567-019-0594-5}
}

@article{Tsirlin2019,
  title = {{Gapless spin-liquid state in the structurally disorder-free triangular antiferromagnet ${\mathrm{NaYbO}}_{2}$}},
  author = {Ding, Lei and Manuel, Pascal and Bachus, Sebastian and Gru\ss{}ler, Franziska and Gegenwart, Philipp and Singleton, John and Johnson, Roger D. and Walker, Helen C. and Adroja, Devashibhai T. and Hillier, Adrian D. and Tsirlin, Alexander A.},
  journal = {Phys. Rev. B},
  volume = {100},
  issue = {14},
  pages = {144432},
  numpages = {9},
  year = {2019},
  month = {Oct},
  publisher = {American Physical Society},
  doi = {10.1103/PhysRevB.100.144432},
  url = {https://link.aps.org/doi/10.1103/PhysRevB.100.144432}
}

@article{Tsirlin2023,
  title = {{Role of alkaline metal in the rare-earth triangular antiferromagnet ${\mathrm{KYbO}}_{2}$}},
  author = {Gru\ss{}ler, Franziska and Hemmida, Mamoun and Bachus, Sebastian and Skourski, Yurii and Krug von Nidda, Hans-Albrecht and Gegenwart, Philipp and Tsirlin, Alexander A.},
  journal = {Phys. Rev. B},
  volume = {107},
  issue = {22},
  pages = {224416},
  numpages = {9},
  year = {2023},
  month = {Jun},
  publisher = {American Physical Society},
  doi = {10.1103/PhysRevB.107.224416},
  url = {https://link.aps.org/doi/10.1103/PhysRevB.107.224416}
}
\let\addcontentsline\oldaddcontentsline

\clearpage


\begin{center}
{\large\bf End Matter} 
\end{center}
\vspace{-8pt}
\section*{Additional numerical results}

\begin{figure}[h]
    \centering
    \includegraphics[trim={1.2cm 0 0 0},clip,width=0.48\linewidth]{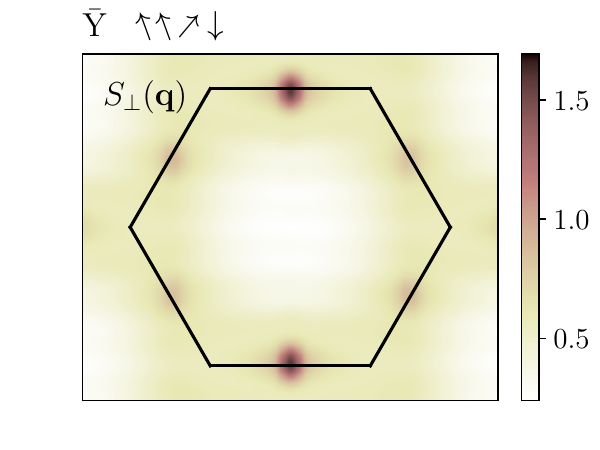}
    \llap{
    \parbox{0pt}{\vspace{-185pt}\hspace{-250pt}\footnotesize{(a)}} 
    }
    \includegraphics[trim={1.2cm 0 0 0},clip,width=0.48\linewidth]{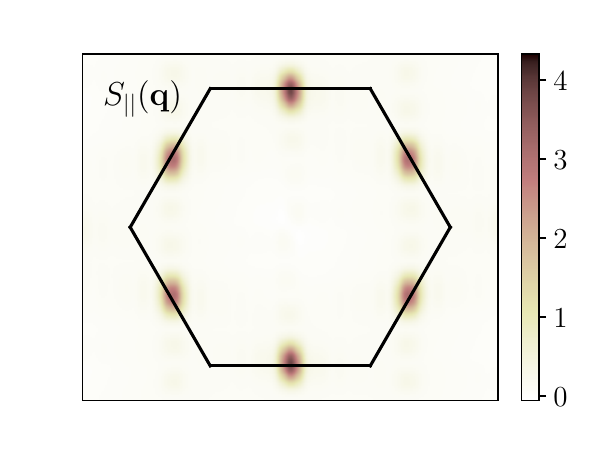} \\
    \includegraphics[trim={1.2cm 0 0 0},clip,width=0.48\linewidth]{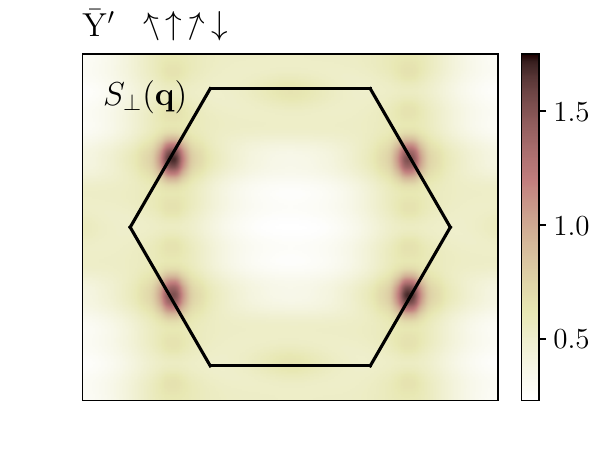}
    \llap{
    \parbox{0pt}{\vspace{-185pt}\hspace{-250pt}\footnotesize{(b)}} 
    }
    \includegraphics[trim={1.2cm 0 0 0},clip,width=0.48\linewidth]{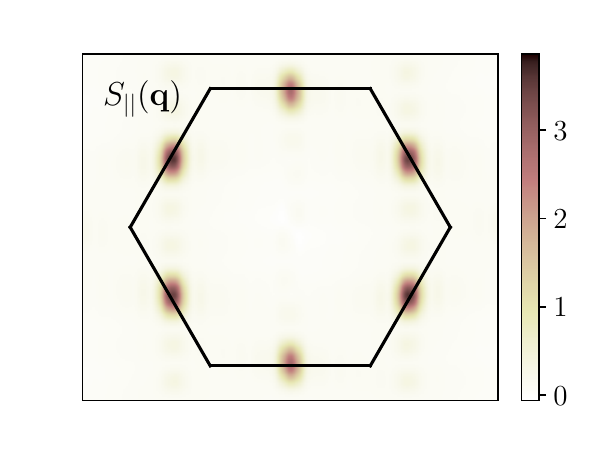} \\
    \caption{Transverse and longitudinal components of the structure factor in the (a) ${\bar{\rm Y}}$ and (b) ${\bar{\rm Y}}'$ states observed in the 4sl-Y region of the phase diagram in Fig.~\ref{fig:PhaseDiagram}.}
    \label{fig:Sq_4sl-Y-like}
\end{figure}

\emph{Four-sublattice Y-like phases.-}
Below the 1/2 plateau we observe a competition between the $\bar{\rm Y}$ state predicted in~\cite{Ye2017,Ye2017b} and a distinct coplanar Y-like state that we denote by $\bar{\rm Y}'$.
The two phases can be distinguished by the transverse component of the structure factor, $S_{\perp}({\bf q})$, as demonstrated in Fig.~\ref{fig:Sq_4sl-Y-like}. While in the $\bar{\rm Y}$ state $S_{\perp}({\bf q})$ shows peaks at all $M$ points (since C$_3$ symmetry is broken in this state the peaks have different intensities, but for all of them it remains finite), in the $\bar{\rm Y}'$ state $S_{\perp}({\bf q})$ shows a 2-Q structure, with vanishing intensity at one of the M points. 
In realistic situations, we expect the quasi-degeneracy between the two phases to be lifted by additional terms that can appear in the Hamiltonian. Furthermore, a cyclic four-spin interaction could drive a transition between these coplanar nematic phases to a 3-Q, C$_3$ invariant cone state.

\begin{figure}[t]
    \centering
    \includegraphics[width=0.48\linewidth]{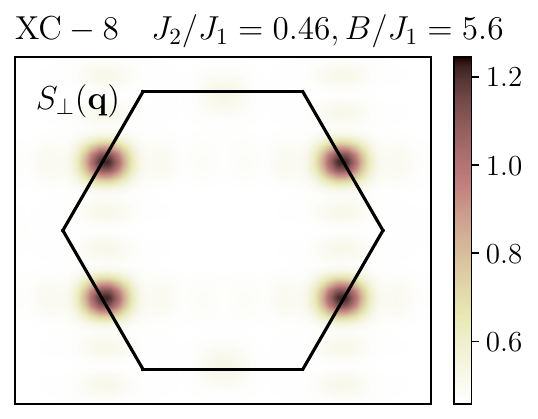}
    \llap{
    \parbox{0pt}{\vspace{-167pt}\hspace{-245pt}\footnotesize{(a)}} 
    }
    \includegraphics[width=0.48\linewidth]{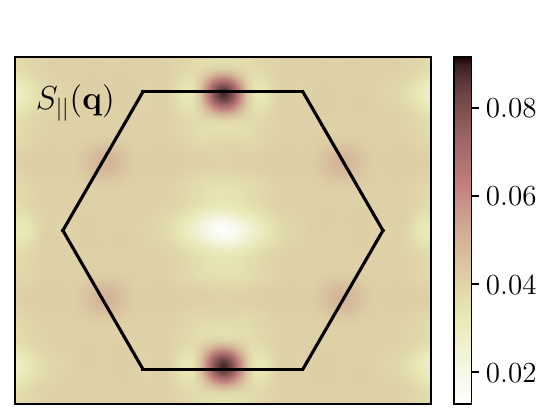} \\
    \includegraphics[width=0.48\linewidth]{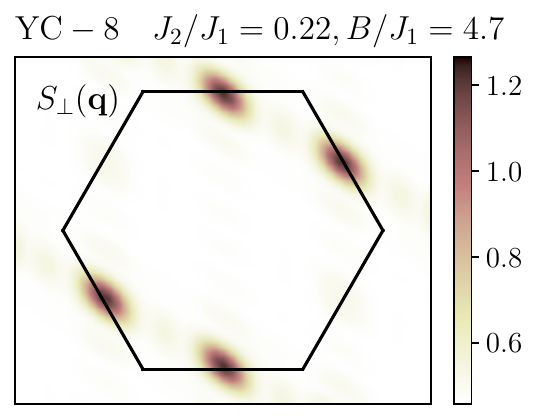}
    \llap{
    \parbox{0pt}{\vspace{-167pt}\hspace{-245pt}\footnotesize{(b)}} 
    }
    \includegraphics[width=0.48\linewidth]{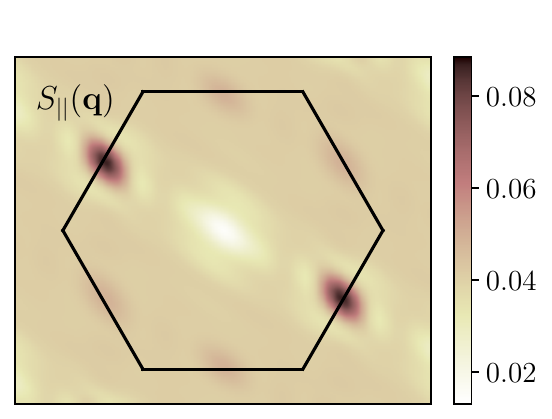}
    \caption{Transverse and longitudinal components of the structure factor in the supersolid stripe state observed for (a) $J_2/J_1>1/3$ on XC-8 cylinders and for (b)  $J_2/J_1<1/3$ on YC-8 cylinders.}
    \label{fig:Sq_supersolid_stripe}
\end{figure}

\emph{High magnetization regime.-}
Just below saturation and away from $J_2/J_1=1/3$, we observe a competition between the canted stripe state and a supersolid stripe state with long-range transverse and longitudinal order. The $J_2$ range in which the supersolid state appears is sensitive to the cylinder geometry: in XC cylinders it occurs for $J_2/J_1>1/3$, while in YC, for $J_2/J_1<1/3$. Fig.~\ref{fig:Sq_supersolid_stripe} shows the corresponding structure factors, where the transverse component exhibits a 2-Q structure — unlike the 1-Q structure of the canted stripe (Fig.~\ref{fig:Sq}(b)) — with the C$_3$ breaking set by the cylinder anisotropy. These observations suggest that the appearance of the supersolid state in our numerics is likely an artifact of the anisotropy introduced by the finite-circumference cylinders. However, it may stabilize in systems with intrinsic lattice anisotropy.

\begin{figure}[h]
    \centering
    \includegraphics[trim={3cm 0 0 0},clip,width=0.48\linewidth]{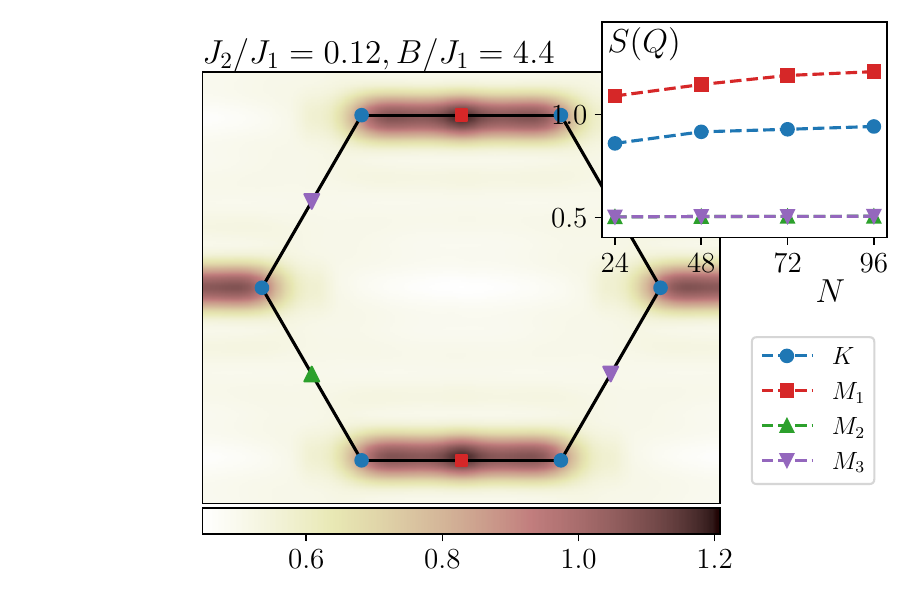}
    \llap{
    \parbox{0pt}{\vspace{-195pt}\hspace{-220pt}\footnotesize{(a)}} 
    }
    \includegraphics[trim={3cm 0 0 0},clip,width=0.48\linewidth]{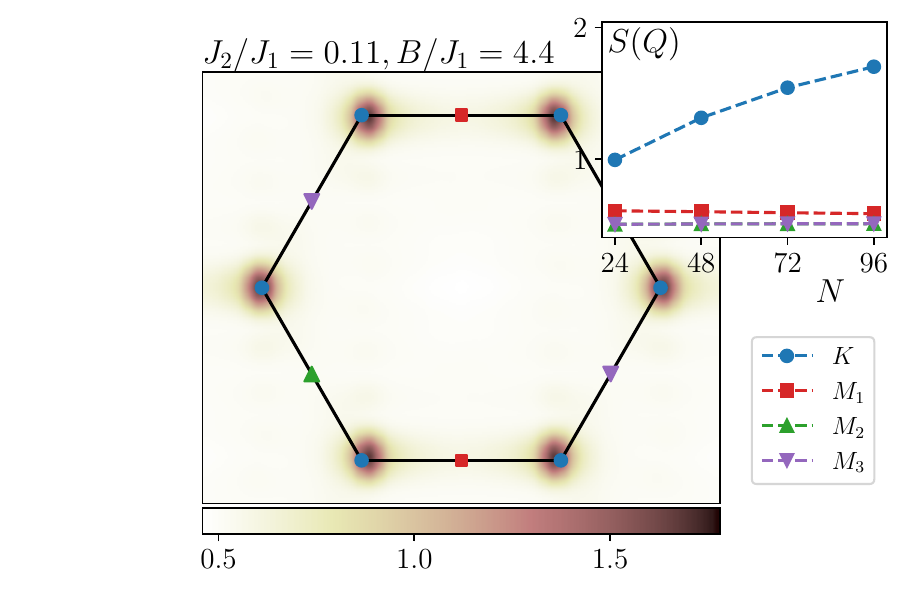}
    \llap{
    \parbox{0pt}{\vspace{-195pt}\hspace{-220pt}\footnotesize{(b)}} 
    }
    \caption{Structure factor in the high magnetization regime ($B/J_1=4.4$) close to the first-order transition line $J_2/J_1=1/8$. (a) For $J_2/J_1=0.12$ the structure factor remains largely featureless, consistent with a quantum-disordered state. (b) For $J_2/J_1=0.11$ clear peaks at the K and K' points can be observed, indicating a long-range-ordered state. }
    \label{fig:Sq_highm}
\end{figure}

In Fig.~\ref{fig:Sq_highm} we plot the structure factor close to saturation in the vicinity of the first-order transition line $J_2/J_1=1/8$, where a quantum-disordered state was predicted by Ref.~\cite{Ye2017}. While in the immediate vicinity of $J_2/J_1=1/8$ the structure factor is indeed featureless, slightly away from this line clear peaks can be observed suggesting the absence of an extended quantum-disordered phase.

\begin{figure}[h]
    \centering
    \includegraphics[trim={1.8cm 0 0 0},clip,width=0.48\linewidth]{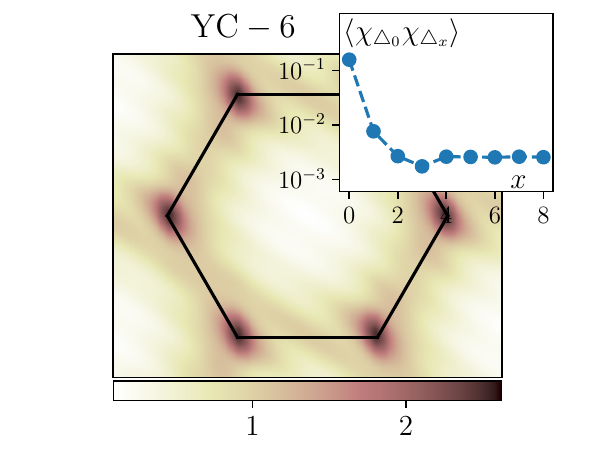}
    \llap{
    \parbox{0pt}{\vspace{-198pt}\hspace{-225pt}\footnotesize{(a)}} 
    }
    \includegraphics[trim={1.8cm 0 0 0},clip,width=0.48\linewidth]{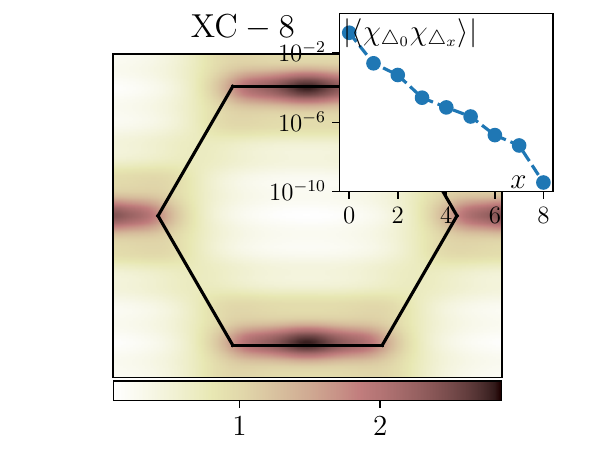}
    \llap{
    \parbox{0pt}{\vspace{-198pt}\hspace{-225pt}\footnotesize{(b)}} 
    }
    \caption{Structure factor for $J_2/J_1=0.12,B/J_1=1$ on (a) YC-6 and (b) XC-8 cylinders. Insets show scalar chirality correlations along a 1D cut, where $x$ is the inter-triangle distance. The cone state on YC-6 cylinders shows peaks at both K and K' points since the wavefunction is restricted to be real-valued in the numerics, resulting in a superposition of states with opposite chirality. The transverse component dominates S({\bf q}) in this case, with longitudinal correlations an order of magnitude smaller. }
    \label{fig:cone}
\end{figure}

\emph{Cone state on $Ny=6$ cylinders and scalar chirality.-}
On cylinders of circumference $N_y=6$,
in the region proximate to the QSL (on the low-$J_2$ side) and partially overlapping it, and at fields in the approximate range $B/J_1 \sim 0.6-1$, we observe the appearance of the non-coplanar cone state characterized by a non-zero scalar chirality. 
The state is clearly favorable in the YC geometry but a strong competition is observed also in the XC geometry.
Using insights from the SW analysis discussed below, we attribute the appearance of the cone state to finite-size effects.
In Fig.~\ref{fig:cone} we plot the structure factor and the scalar chirality correlations obtained at the same point in the phase diagram ($J_2/J_1=0.12,B/J_1=1$) on a YC-6 cylinder in (a) and an XC-8 cylinder in (b). The former hosts the cone state with long-range scalar chirality correlations, while the latter shows a disordered state with rapidly decaying correlations.

\section*{Spin-wave calculations}

\emph{LSWT analysis of magnetization and susceptibility at low fields.-}
In describing three-sublattice states, we closely follow Ref.~\onlinecite{Chernyshev2022}.
Our LSW Hamiltonian is given by Eq. (22) of the reference above (with $b=0$ in Eq.(25,26)). The order parameter for the sublattice $\alpha = A,B,C$ is, in notations of \cite{Chernyshev2022} and using Eq.(28), then given by $\langle S^z_{\bf r}\rangle = S - \langle a^\dagger_{\alpha,{\bf r}} a_{\alpha,{\bf r}}\rangle = N_c^{-1}\sum_{{\bf q}\in {\rm MBZ}} \sum_{\gamma=1,2,3} |V^{(\gamma)}_{\alpha, {\bf q}}|^2$,
where MBZ denotes the magnetic BZ. 
The required ${\bf q}$-integration
can only be done numerically, using Gaussian quadrature technique, and is time-consuming. We find that reformulating the matrix diagonalization problem as the matrix equation for boson Green's functions along the lines of \cite{Krivoruchko1981,Maksimov2019} simplifies it significantly. 

To find the zero-field uniform susceptibility, we write $6\times6$ matrix (Eq.(23) in \cite{Chernyshev2022}) $\hat{\bf H}_{\bf q} = \sum_{\ell=0}^2 \hat{\bf H}_{\bf q}^{(\ell)} h^\ell$ and diagonalize $\hat{\bf H}_{\bf q}^{(\ell=0)}$, which describes the $120^\circ$ state, first. Subsequently, we treat $\hat{\bf H}_{\bf q}^{(\ell=1,2)}$ as a perturbation and evaluate its contribution to the ground state energy of the Y state to $h^2$ order and read off the susceptibility from there. We obtain
\bea
\label{eq2}
\chi_{\rm Y} = &&\frac{1}{9J_1}\Big[1-\frac{1}{2 S N}\sum_{{\bf k} \in {\rm BZ}} \frac{(a_{\bf k} - b_{\bf k})\bar{\gamma}_{\bf k}}{E_{\bf k}} +\\
&&+\frac{3}{2 S N}\sum_{{\bf k} \in {\rm BZ}} \frac{(a_{{\bf k}+{\bf Q}} - b_{{\bf k}+{\bf Q}})(a_{{\bf k}-{\bf Q}} - b_{{\bf k}-{\bf Q}})\bar{\gamma}_{\bf k}^2}{ E_{{\bf k}+{\bf Q}} E_{{\bf k}-{\bf Q}}(E_{{\bf k}+{\bf Q}} + E_{{\bf k}-{\bf Q}})}\Big]
\nonumber
\eea
Here, $j_2 = J_2/J_1$, $a_{\bf k} = 1 + \bar{\gamma}_{\bf k}/2 - 2 j_2 (1-\gamma^{(2)}_{\bf k})$, $b_{\bf k} = -3 \bar{\gamma}_{\bf k}/2$, $E_{\bf k} = \sqrt{a_{\bf k}^2 - b_{\bf k}^2}$ is the magnon dispersion in the $120^\circ$ state, and ${\bf Q} = (4\pi/3,0)$ \cite{Chernyshev2022}. 
Notice that here the sum is over the full BZ of the triangular lattice. Eq.\eqref{eq2} generalizes the previous $j_2=0$ result \cite{Chubukov_1991,Chubukov1994}.  

To compare with the cone (umbrella) state, we also calculated its zero-field susceptibility following \cite{Veillette2005},
\be
\label{eq3}
\chi_{\rm cone} = \frac{1}{9J_1}\Big[1+\frac{1}{2 S N}\sum_{{\bf k} \in {\rm BZ}} \frac{(a_{\bf k} + b_{\bf k})\bar{\gamma}_{\bf k}}{E_{\bf k}}\Big].
\ee
Equations \eqref{eq2} and \eqref{eq3} show that while classically Y and cone states are degenerate, i.e. $\chi_{\rm Y} = \chi_{\rm cone}$ for $S=\infty$, quantum corrections select the Y state at finite $S$ for all values of $0 \leq j_2 \leq 0.125$.

Analysis of the two-sublattice stripe phase is simpler. We transform the stripe-state spin-wave Hamiltonian to the rotating basis with ${\bf Q} \to {\bf M}'= (0, 2\pi/\sqrt{3})$ \cite{Mourigal2010}. 
This reduces the problem to one involving a single spin-wave boson species, 
\bea
\label{eq4}
&&\chi_{\rm stripe} = \frac{1}{8 J_1(1+j_2)}\Big[1 - \frac{1}{2 S (1+j_2) N} \sum_{{\bf k} \in {\rm BZ}} g_{\bf k} \times \nonumber\\
&&\times \sqrt{\frac{1+ j_2 + \cos[k_x]+j_2 \cos[\sqrt{3} k_y] +2 g_{\bf k}}{1+ j_2 + \cos[k_x]+j_2 \cos[\sqrt{3} k_y]-2 g_{\bf k}}} \, \Big], 
\eea
where $g_{\bf k} = (\cos[k_x/2] + j_2 \cos[3k_x/2]) \cos[\sqrt{3} k_y/2]$.

It is instructive to repeat this analysis in the cylinder geometry used in DMRG. Due to the finite circumference, momentum perpendicular to the cylinder axis is quantized, and thus the 2D integration of ${\bf k}$ has to be replaced by a 1D integration over $k_x$ and summation over discrete $k_y$ values. 
Unlike the 2D case, we find $\chi_{\rm cone} > \chi_{\rm Y}$ for cylinders with $N_y < 24$. This is consistent with the tendency towards the cone state observed on $N_y=6$ cylinders, and  stresses the need for a cautious interpretation of DMRG data even for relatively wide cylinders.

\emph{Self-consistent analysis for the magnetization plateaux.-}
To determine plateaux boundaries we fix $J_2/J_1$ and track the closing of the single-magnon gap as the field is scanned around the classical stability value $B_{\text{cl}}=3J_1S$ ($4(J_1+J_2)S$) for the UUD (UUUD) plateau. At the level of LSWT, the single-magnon spectrum is gapless due to accidental degeneracy.  
This degeneracy is lifted by one-loop (OL) corrections \cite{AliceaJ2009}.
The plateau states contain no cubic interaction vertices,
so the corrections arise entirely from normal ordering (NO) the quartic terms, yielding the renormalized quadratic Hamiltonian $\mathcal{H}^{(2)}=\mathcal{H}^{(2)}_{\text{LSWT}}+\mathcal{H}^{(2)}_{\text{NO}}$. The matrix elements of $\mathcal{H}^{(2)}_{\text{NO}}$ are given by linear combinations of vacuum expectation values of boson bilinears $\langle a_i^{\dagger} a_j \rangle$ and $\langle a_i a_j \rangle$. In the OL scheme (see \cite{AliceaJ2009,Starykh2014} for details), these averages are evaluated in the LSWT vacuum. 
In contrast, the self-consistent scheme updates \emph{all} matrix elements of $\mathcal{H}^{(2)}$, recomputes the bilinear expectation values in the corresponding vacuum, and iterates the procedure until self-consistency is reached.

\clearpage

\onecolumngrid
\setcounter{equation}{0}
\setcounter{figure}{0}
\renewcommand{\theequation}{S\arabic{equation}}
\renewcommand{\thefigure}{S\arabic{figure}}

\renewcommand{\thesection}{S\arabic{section}}
\renewcommand{\thesubsection}{\thesection.\arabic{subsection}}
\renewcommand{\thesubsubsection}{\thesubsection.\arabic{subsubsection}}

\begin{center}
{\Large\bfseries Supplementary Material}    
\end{center}

\tableofcontents

\section{Linear spin-wave theory analysis of sublattice magnetizations in the Y state} 
In describing three-sublattice states, we closely follow Ref.~\cite{Chernyshev2022}.
Our LSW Hamiltonian is given by Eq.~(22) of this reference above with $b=0$ in Eqs.~(25) and (26). More details are provided below.

For a spin-wave expansion,  the laboratory reference frame  $\{x_0,y_0,z_0\}$ in 
is rotated to the 
{\it local} reference frame  $\{x,y,z\}$ on each site, so that  the $z$ axis is 
along the direction dictated by a classical spin configuration obtained.
For the coplanar Y and V states in  Fig.~4 of Ref.~\cite{Chernyshev2022}, such a transformation is  a simple rotation 
in the $x_0$--$z_0$  plane, such that $S_\alpha^{y_0}\!=\!S_\alpha^y$ and 
\begin{align}
\label{Sxyz}
&S^{x_0}_\alpha=S^x_\alpha\cos\widetilde{\alpha} -S^z_\alpha\sin\widetilde{\alpha}, \\ 
&S^{z_0}_\alpha=S^z_\alpha\cos\widetilde{\alpha} +S^x_\alpha\sin\widetilde{\alpha}, \nonumber
\end{align}
where $\alpha$  and $\widetilde{\alpha}$ are, respectively, sublattice index and corresponding  spin angles in Fig.~4(b) of Ref.~\cite{Chernyshev2022}.

This is followed by the standard Holstein-Primakoff bosonization of spins
in the rotated reference frame,
$S_i^z\!=\! S- a^\dagger_{i}a^{\phantom{\dag}}_{i}$ and 
$S_{i}^-\!=\!a_{i}^\dag\sqrt{2S}$.

In the three-sublattice state, the site coordinate $i$ is replaced by the unit cell coordinate
 $\ell$ and sublattice index $\alpha$:
$i\!\rightarrow\!\{\alpha,\ell\}$. 
As a result, the Holstein-Primakoff boson operators 
are split into three species 
$a^{(\dag)}_{\alpha,\ell}\!=\!\{a_\ell^{(\dag)},b_\ell^{(\dag)},c_\ell^{(\dag)}\}$  corresponding to 
$\alpha\!=\!\{{\rm A},{\rm B},{\rm C}\}$ sublattices.
The Fourier transformation for them is
\begin{equation}
\label{FT3}
a^{\phantom{\dag}}_{\alpha,\ell} = \frac{1}{\sqrt{N_c}} \sum_{\bf q}  \,
a^{\phantom{\dag}}_{\alpha,\bf q} \, e^{-i{\bf q}\cdot{\bf r}_{\alpha,\ell}},
\end{equation}
where ${\bf r}_{\alpha,\ell}\! =\! {\bf R}_\ell\! +\! \bm{\rho}_\alpha$ and $N_c\!=\!N/3$ 
is the number of unit cells.
The  sublattice coordinates within the unit cell can be chosen as  
$\bm{\rho}_A\! =\! 0$, $\bm{\rho}_B\! =\! -\bm{\delta}_2$, and
$\bm{\rho}_C \!=\! \bm{\delta}_3$, see Fig.~3 of Ref.~\cite{Chernyshev2022}.

Using these boson species and their Fourier transforms, 
the LSWT Hamiltonian for an arbitrary coplanar three-sublattice state reads as
\begin{align}
\label{LSWT_H}
\hat{\cal H}^{(2)} =\frac{3J_1S}{2}\sum_{\bf q}\hat{\bf x}_{\bf q}^\dagger 
\hat{\bf H}_{\bf q}\hat{\bf x}_{\bf q}^{\phantom{\dagger}}\, ,
\end{align}
where $\hat{\bf x}^\dag_{\bf q}\!=\!\left( a^\dag_{\bf q}, b^\dag_{\bf q},c^\dag_{\bf q}, 
a^{\phantom \dag}_{-\bf q},b^{\phantom \dag}_{-\bf q},c^{\phantom \dag}_{-\bf q}\right)$ 
and $\hat{\bf H}_{\bf q}$ is a matrix
\begin{eqnarray}
\label{LSWTmatrix}
\hat{\bf H}_{\bf q}= 
\left( \begin{array}{cc} 
\hat{\bf A}^{\phantom \dagger}_{\bf q} &  \hat{\bf B}^{\phantom \dagger}_{\bf q}\\[0.5ex] 
\hat{\bf B}^\dagger_{\bf q}  & \hat{\bf A}^*_{-\bf q}
\end{array}\right), 
\end{eqnarray}
with the  $3\times 3$ matrices
$\hat{\bf A}^{\phantom \dag}_{\bf q}$ and $\hat{\bf B}^{\phantom \dag}_{\bf q}$  
\begin{eqnarray}
\label{ABany}
\hat{\bf A}^{\phantom \dag}_{\bf q}=
\left( \begin{array}{ccc} 
A_{\bf q} &  D_{\bf q} &  E^*_{\bf q}\\
D^*_{\bf q}  & B_{\bf q} &  F_{\bf q} \\
E_{\bf q}  & F^*_{\bf q} &  C_{\bf q}
\end{array}\right), \ \ \  
\hat{\bf B}^{\phantom \dag}_{\bf q}=
\left( \begin{array}{ccc} 
G &  J_{\bf q} & K^*_{\bf q} \\
J^*_{\bf q}  & H & L_{\bf q}\\
K_{\bf q}  & L^*_{\bf q} &  I
\end{array}\right). \ \ \ \ \ \ 
\end{eqnarray}
The elements of the $\hat{\bf A}^{\phantom \dag}_{\bf q}$ matrix are given by
\begin{align}
A_{\bf q}=&\, h\cos\beta-\cos\widetilde{\alpha}_{AB}-\cos\widetilde{\alpha}_{AC}
-2j_2\big(1-\gamma^{(2)}_{\bf q}\big),
\nonumber\\
B_{\bf q}=&\, h\cos\alpha_1-\cos\widetilde{\alpha}_{AB}-\cos\widetilde{\alpha}_{BC}
-2j_2\big(1-\gamma^{(2)}_{\bf q}\big),
\nonumber\\
C_{\bf q}=&\, h\cos\alpha_2-\cos\widetilde{\alpha}_{BC}-\cos\widetilde{\alpha}_{AC}
-2j_2\big(1-\gamma^{(2)}_{\bf q}\big),\nonumber\\
\label{ABCD_any}
D_{\bf q}=&\, \gamma_{\bf q}\cos^2(\widetilde{\alpha}_{AB}/2),
\nonumber\\
E_{\bf q}=&\, \gamma_{\bf q}\cos^2(\widetilde{\alpha}_{AC}/2),
\nonumber\\
F_{\bf q}=&\, \gamma_{\bf q}\cos^2(\widetilde{\alpha}_{BC}/2),
\end{align}
and of the  $\hat{\bf B}^{\phantom \dag}_{\bf q}$ matrix, respectively,
\begin{align}
G &= H = I = 0,\nonumber\\
J_{\bf q}=&  -\gamma_{\bf q}\sin^2(\widetilde{\alpha}_{AB}/2),
\nonumber\\
K_{\bf q}=&  -\gamma_{\bf q}\sin^2(\widetilde{\alpha}_{AC}/2),
\nonumber\\
L_{\bf q}=&  -\gamma_{\bf q}\sin^2(\widetilde{\alpha}_{BC}/2),
\label{GHI_any}
\end{align}
where  $h\!=\!g\mu_BH/3J_1S$, $j_2\!=\!J_2/J_1$, 
$\widetilde{\alpha}_{AB}\!=\!\beta-\alpha_1$, $\widetilde{\alpha}_{AC}\!=\!\beta-\alpha_2$,
$\widetilde{\alpha}_{BC}\!=\!\alpha_1-\alpha_2$, and 
\begin{align}
\label{gammas}
\gamma_{\bf q}=\frac{1}{3}\sum_{\alpha}  e^{i{\bf q} \cdot\bm{\delta}_\alpha},\ \ \ 
\gamma^{(2)}_{\bf q}=\frac{1}{3}\sum_{\alpha}  \cos {\bf q}{\cdot}\bm{a}_\alpha\, ,
\end{align}
with the first- and second-neighbor translation vectors
$\bm{\delta}_1 \!=\! (1,0)a$,
$\bm{\delta}_2 \!=\! (-1, \sqrt{3})a/2$, 
$\bm{\delta}_3 \!=\!- (1, \sqrt{3})a/2$, and 
$\bm{a}_1 \!=\! (3, -\sqrt{3})a/2$,
$\bm{a}_3 \!=\! (0, \sqrt{3})a$, 
$\bm{a}_2 \!=\! -(3, \sqrt{3})a/2$, respectively, see Fig.~3 of Ref.~\cite{Chernyshev2022}; $a$ is the lattice constant. The equilibrium angles of the Y state are: $\beta = \pi$, $\alpha_1 = - \alpha_2 = \arccos[(1+h)/2]$.

The next step is to find the coefficients  $U$ and $V$ of the generalized Bogolyubov transformation from the Holstein-Primakoff 
bosons to the quasiparticle eigenmodes
\begin{align}
\label{Bogo}
a^{\phantom{\dag}}_{\alpha,{\bf q}}=\sum_\gamma\left(U_{\alpha,{\bf q}}^{({\gamma})}\,
A^{\phantom{\dag}}_{\gamma,{\bf q}}+V_{\alpha,{\bf q}}^{({\gamma})}\,
A^{\dag}_{\gamma,-{\bf q}}\right),
\end{align}
with the  quasiparticle operators $A_{\gamma,{\bf q}}\!=\!\left\{A_{\bf q},B_{\bf q},C_{\bf q}\right\}$. 

In matrix form, the transformation \eqref{Bogo} is
\begin{eqnarray}
\label{UVtrans}
\hat{\bf x}_{\bf q}\!=\!\left(\! \begin{array}{c} 
\hat{\bf a}^{\phantom \dagger}_{\bf q} \\[0.5ex] 
\hat{\bf a}^\dagger_{-\bf q}  
\end{array}\!\right) \!=\! 
\left( \!\begin{array}{cc} 
\hat{\bf U}_{\bf q} &  \hat{\bf V}_{\bf q}\\[0.5ex] 
\hat{\bf V}^*_{-\bf q} & \hat{\bf U}^*_{-\bf q}
\end{array}\!\right)\!
\left( \!\begin{array}{c} 
\hat{\bf \cal A}^{\phantom \dagger}_{\bf q} \\[0.5ex] 
\hat{\bf \cal A}^\dagger_{-\bf q}  
\end{array}\!\right)\!= \hat{\bf S}_{\bf q} \cdot\hat{\bf z}_{\bf q}, \ \ \
\end{eqnarray}
where vectors $\hat{\bf a}_{\bf q}\!=\!\big[ a_{\bf q}, b_{\bf q},c_{\bf q}\big]^T$,  
$\hat{\bf a}^\dag_{-\bf q}\!=\!\big[ a^\dag_{-\bf q}, b^\dag_{-\bf q},c^\dag_{-\bf q}\big]^T$ and 
$\hat{\bf \cal A}_{\bf q}\!=\!\big[ A_{\bf q}, B_{\bf q},C_{\bf q}\big]^T$,
$\hat{\bf \cal A}^\dag_{-\bf q}\!=\!\big[ A^\dag_{-\bf q}, B^\dag_{-\bf q},C^\dag_{-\bf q}\big]^T$ are
introduced. It follows that the transformation matrix $\hat{\bf S}_{\bf q}$ diagonalizes  
$\hat{\bf g} \hat{\bf H}_{\bf q}$ in \eqref{LSWTmatrix}, where the ``commutation" matrix $\hat{\bf g}$ is defined by $\hat{\bf g} = {\rm Diag}(1,1,1,-1,-1,-1)$, as explained in Ref.~\cite{Chernyshev2022}.
Thus, the $U_{\alpha}^{({\gamma})}$ and $V_{\alpha}^{({\gamma})}$ parameters
can be extracted as the elements of the properly normalized eigenvectors of $\hat{\bf g} \hat{\bf H}_{\bf q}$
from a diagonalization procedure.

It is useful to observe that $\hat{\bf A}^*_{-\bf q}\!=\!\hat{\bf A}_{\bf q}$ and 
$\hat{\bf B}^\dag_{\bf q}\!=\!\hat{\bf B}_{\bf q}$ in \eqref{LSWTmatrix} as 
their off-diagonal matrix elements \eqref{ABCD_any}, \eqref{GHI_any} are simply proportional to 
the complex hopping amplitude $\gamma^*_{-\bf q}\!=\!\gamma_{\bf q}$ in Eq.~\eqref{gammas}.
As a result, all eigenenergies of $\hat{\bf g} \hat{\bf H}_{\bf q}$ 
are reciprocal, $\omega_{\gamma -{\bf q}}\!=\!\omega_{\gamma{\bf q}}$,
and $\{\hat{\bf V}^*_{-\bf q},\hat{\bf U}^*_{-\bf q}\}\!=\!\{\hat{\bf V}_{\bf q},\hat{\bf U}_{\bf q}\}$ 
in \eqref{UVtrans}.

At this point we observe that the order parameter for the sublattice $\alpha = A,B,C$ is given, for $S=1/2$, by 
\begin{align}
\label{Sz}
\langle S^z_{\bf r}\rangle = \frac{1}{2} - \langle a^\dagger_{\alpha,{\bf r}} a_{\alpha,{\bf r}}\rangle = \frac{1}{2} - N_c^{-1}\sum_{{\bf q}\in {\rm MBZ}} \sum_{\gamma=1,2,3} |V^{(\gamma)}_{\alpha, {\bf q}}|^2,
\end{align}
where MBZ denotes the magnetic BZ. Importantly, we do not need $U, V$ coefficients per se, only the specific combination $\sum_{\gamma=1,2,3} |V^{(\gamma)}_{\alpha, {\bf q}}|^2$.

To evaluate \eqref{Sz}, we reformulate the matrix diagonalization problem as the matrix equation for boson Green's functions along the lines of \cite{Krivoruchko1981,Maksimov2019}, simplifying it significantly. Specifically, we define retarded Green's functions 
\begin{align}
    G_{j1}^{-+}(k,t) = -i \theta(t) \langle [\hat{a}_{jk}(t), \hat{a}_{1k}^\dagger(0)]\rangle, \, G_{j1}^{++}(k,t) = -i \theta(t) \langle [\hat{a}_{j-k}^\dagger(t), \hat{a}_{1k}^\dagger(0)]\rangle, 
\end{align}
where $\hat{a}_{jk}$ (or its conjugate) represents the $j$-th element of the operator vector $\hat{\bf a}_{\bf k}$ (or its conjugate), with $j=1,2,3$. These obey the equation of motion 
\begin{align}
    -i \partial_t G_{j1}^{-+}(k,t) = - \delta(t) \langle [\hat{a}_{jk}(0), \hat{a}_{1k}^\dagger(0)]\rangle - i \theta(t) \langle [[\hat{\cal H}^{(2)},\hat{a}_{jk}(t)], \hat{a}_{1k}^\dagger(0)]\rangle .
\end{align}
The first term on the right-hand side is just $\delta_{j1}$, of course. Equation of motion for $G_{j1}^{++}(k,t)$ does not have a similar term because the commutator of two creation (or annihilation) operators is zero. Using the explict form of the Hamiltonian \eqref{LSWT_H} and Fourier transforming as $G(t) = \int \frac{d\omega}{2\pi} e^{-i \omega t} G(\omega)$, one arrives at the matrix equation 
\begin{equation}
\big(\hat{\bf H}_{\bf k} - \omega \hat{\bf g}\big)_{ji} 	
    \begin{pmatrix} 
    G^{-+}_{i 1} \\ 
    G^{++}_{i 1}
    \end{pmatrix}
    = 
    \begin{pmatrix} 
    -1 \\ 
    {\bf 0}_5
    \end{pmatrix}_{j 1}
\end{equation}
where the Hamiltonian matrix on the left is defined in Eq.~\eqref{LSWTmatrix}. 
The vector $(-1, {\bf 0}_5)^{\rm T}$ on the right-hand side is symbolic notation for $(-1,0,0,0,0,0)^{\rm T}$, while that on the left-hand side is made of the two length-$3$ vectors of the Green's functions defined above.

This equation can now be solved by Cramer's rule: the matrix equation $\hat{A} {\bf x} = {\bf b}$, where $\hat{A}$ is an $n\times n$ matrix with a nonzero determinant, is solved by $x_i = {\rm det}(\hat{A}_i)/{\rm det}(\hat{A})$, where $\hat{A}_i$
 is the matrix formed by replacing the $i$-th column of $\hat{A}$ by the column vector ${\bf b}$.

In our case, the determinant in the denominator is just ${\rm det}(\hat{\bf H}_{\bf k} - \omega \hat{\bf g}) = \Pi_{n=1}^3(\epsilon_n(k) - \omega)(\epsilon_n(k) + \omega)$, in terms of magnon eigenvalues $\epsilon_n(k)$. Therefore, the Green's function $G_{11}^{-+}(k,\omega)$ has poles at $\omega = \pm \epsilon_n(k)$. In particular, the residue of the pole at $\omega = \epsilon_n(k)$ is given by 
\begin{align}
\label{U-cramer}
    |U_{1,{\bf k}}^{(n)}|^2 = -\frac{{\rm det}(\hat{\bf H}_{\bf k} - \epsilon_n(k) \hat{\bf g})_1}{2\epsilon_n(k) \Pi_{m\neq n}(\epsilon_n(k)^2-\epsilon_m(k)^2)},
\end{align}
where the matrix $(\hat{\bf H}_{\bf k} - \epsilon_n(k) \hat{\bf g})_1$ in the numerator is the analogue of $\hat{A}_1$ of the Cramer's rule.

The identification of the residue with $|U_{1,{\bf k}}^{(n)}|^2$ follows from applying eigenmode expansion \eqref{Bogo} to $G_{11}^{-+}(k,\omega)$, which results in 
\begin{align}
    G_{11}^{-+}(k,\omega) = \sum_{n=1}^3 \Big(\frac{|U_{1,{\bf k}}^{(n)}|^2}{\omega - \epsilon_n(k)+i0} - \frac{|V_{1,{\bf k}}^{(n)}|^2}{\omega + \epsilon_n(-k)+i0} \Big).
\end{align}
Matching the pole of this expression at $\omega = \epsilon_n(k)$ to the one obtained from Cramer's rule leads to the identification \eqref{U-cramer}.

Finally, using the standard $U-V$ identity
\begin{align}
\label{UVnorm}
\sum_\gamma\left(\big|U_{\alpha,{\bf q}}^{({\gamma})}\big|^2-
\big|V_{\alpha,{\bf q}}^{({\gamma})}\big|^2\right)=1,
\end{align}
gives us an algebraic way to evaluate $\sum_{\gamma=1}^3\big|V_{\alpha,{\bf q}}^{({\gamma})}\big|^2$ at momentum ${\bf q}$ in the magnetic BZ in Eq.~\eqref{Sz}. The described manipulations illustrate the case of the A-sublattice, when $\alpha =1$, resulting in $\langle S^z_A\rangle$. Obviously, its extension to B/C sublattices requires one to analyze Green's functions $G_{j \,2/3}^{\pm +}$ in a similar way.

The remaining sum over ${\bf q}$ is done numerically, using the Gaussian quadrature technique. All of the described matrix manipulations are implemented in the Mathematica program. The results of the calculation are illustrated by the dashed red curves with up/down triangle markers in 
Fig.~4(b) of the main text. 

\section{Linear spin-wave theory analysis of sublattice magnetizations in the stripe state}
Next, we present a single-boson formulation of the stripe phase. This is based on the sublattice-dependent rotation of spins from the laboratory frame $(x_0,y_0,z_0)$ to the local one $(x,y,z)$. For the spin operator at the site ${\bf r}$,
\begin{align}
    S^{x_0}_{\bf r} = \sin\theta S^x_{\bf r} + \cos\theta e^{i M_1 \cdot {\bf r}} S^z_{\bf r} , S^{y_0}_{\bf r} = S^y_{\bf r} , S^{z_0}_{\bf r} = - \cos\theta e^{i M_1 \cdot {\bf r}} S^x_{\bf r} + \sin\theta S^z_{\bf r}.
\end{align}
Here $M_1 = (0,2\pi/\sqrt{3})$, and the spin-wave expansion in the local frame is $S^z = S - a^+ a, S^x = \sqrt{S/2}(a +a^+), S^y = -i\sqrt{S/2}(a-a^+)$. By requiring the linear in magnons $a, a^+$ terms to vanish, the optimal rotation angle is found to be $\sin\theta = B/(8S(J_1 + J_2))$. The harmonic Hamiltonian of the spin waves in the stripe phase is found to be
\begin{eqnarray}
    H^{(2)}_{\rm stripe} &=& S \sum_{{\bf k} \in {\rm BZ}} A_{\bf k} a^+_{\bf k} a_{\bf k} - \frac{1}{2}B_{\bf k} (a_{\bf k} a_{-{\bf k}} + {\rm h.c.}), \\
    A_{\bf k} &=& 2J_1 \Big[1 + \cos{\bf k}\cdot {\bm \delta}_1 + \sin^2\theta (\cos{\bf k}\cdot {\bm \delta}_2 + \cos{\bf k}\cdot {\bm \delta}_3)\Big]+2J_2 \Big[1 + \cos{\bf k}\cdot {\bf e}_2 + \sin^2\theta (\cos{\bf k}\cdot {\bf e}_1 + \cos{\bf k}\cdot {\bf e}_3)\Big], \nonumber\\
    B_{\bf k} &=& \cos^2\theta \Big[2J_1 (\cos{\bf k}\cdot {\bm \delta}_2 + \cos{\bf k}\cdot {\bm \delta}_3) + 2J_2 (\cos{\bf k}\cdot {\bf e}_1 + \cos{\bf k}\cdot {\bf e}_3)\Big]\nonumber
\end{eqnarray}
As usual, the magnon dispersion is then $\omega_{\bf k} = \sqrt{A_{\bf k}^2 - B_{\bf k}^2}$, while the ordered moment 
\begin{equation}
    \langle S^z \rangle = \frac{1}{2} - \frac{1}{N} \sum_{{\bf k} \in {\rm BZ}} \Big(\frac{A_{\bf k}}{\omega_{\bf k}} - 1\Big).
\end{equation}
The last expression is used to derive the spin-wave boundary of the stripe phase in Fig.~4(b) 
(dashed red line with square markers). The magnetization and susceptibility, Eq.~(4) of the main text/End Matter, are obtained by subsequent differentiating of the ground state energy $E_{\rm gs} = H^{(0)}_{\rm stripe} + \frac{1}{2} \sum_{\bf k} (\omega_{\bf k} - A_{\bf k})$ with respect to $B$. Here, the classical energy $H^{(0)}_{\rm stripe} = -S^2 (J_1 + J_2) - B^2/(16(J_1+J_2))$.

\section{Semiclassical spin-wave theory analysis for the magnetization plateaux}
We first note that the UUD (UUUD) configuration is a classical ground state only at the fine-tuned magnetic field
$B_{\mathrm{cl}}=3J_1S$ [$B_{\mathrm{cl}}=4(J_1+J_2)S$]. 
Following Refs.~\cite{AliceaJ2009,KamiyaY2018a}, all spin-wave calculations within the plateau are therefore initialized at this classical field $B_{\mathrm{cl}}$. 
In the following, we present the calculation for the UUD order as an illustrative example. 
The results for both the UUD and UUUD plateaux can be reproduced using the numerical code provided in the accompanying repository~\cite{Zhang2026TriangularPlateaux}, which implements the derivations below within the Sunny platform~\cite{Sunny_software,Sunny_paper}.

In the UUD phase, the LSWT Hamiltonian can be written in the Bogoliubov form
\begin{equation}
    \mathcal{H}_{\mathrm{LSWT}}
    =
    \frac{1}{2}\sum_{\bm{q}\in\mathrm{MBZ}}
    \begin{pmatrix}
        (\bm{a}_{\bm{q}}^{\dagger})^T & (\bm{a}_{-\bm{q}})^T
    \end{pmatrix}
    H_{0}^{(2)}(\bm{q})
    \begin{pmatrix}
        \bm{a}_{\bm{q}} \\
        \bm{a}_{-\bm{q}}^{\dagger}
    \end{pmatrix},
    \qquad
    H_{0}^{(2)}(\bm{q})
    \equiv
    \begin{pmatrix}
        H^{0}_{11}(\bm{q}) & H^{0}_{12}(\bm{q}) \\
        H^{0}_{21}(\bm{q}) & H^{0}_{22}(\bm{q})
    \end{pmatrix}.
    \label{eq:uud_lswt_ham}
\end{equation}
Here the momentum summation is restricted to the magnetic Brillouin zone (MBZ), and
\begin{equation}
    \bm{a}_{\bm{q}}
    =
    \begin{pmatrix}
       a_{1,\bm{q}} \\
       a_{2,\bm{q}} \\
       a_{3,\bm{q}}
    \end{pmatrix},
    \qquad
    \bm{a}_{-\bm{q}}^{\dagger}
    =
    \begin{pmatrix}
       a_{1,-\bm{q}}^{\dagger} \\
       a_{2,-\bm{q}}^{\dagger} \\
       a_{3,-\bm{q}}^{\dagger}
    \end{pmatrix}
\end{equation}
are the Holstein-Primakoff (H-P) boson spinors associated with the three magnetic sublattices of the UUD state. The block matrices are given by
\begin{equation}
    H_{11}^{0}(\bm{q})
    =
    S
    \begin{pmatrix}
        BS^{-1}+6J_2(\gamma_{\bm{q}}^{(2)}-1)
        &
        3J_1 \gamma_{\bm{q}}^{(1)}
        &
        0
        \\
        3J_1 \gamma_{-\bm{q}}^{(1)}
        &
        BS^{-1}+6J_2(\gamma_{\bm{q}}^{(2)}-1)
        &
        0
        \\
        0
        &
        0
        &
        6J_1-BS^{-1}+6J_2(\gamma_{\bm{q}}^{(2)}-1)
    \end{pmatrix},
\end{equation}
and
\begin{equation}
    H_{12}^{0}(\bm{q})
    =
    S
    \begin{pmatrix}
        0
        &
        0
        &
        -3J_1\gamma_{-\bm{q}}^{(1)}
        \\
        0
        &
        0
        &
        -3J_1\gamma_{\bm{q}}^{(1)}
        \\
        -3J_1\gamma_{\bm{q}}^{(1)}
        &
        -3J_1\gamma_{-\bm{q}}^{(1)}
        &
        0
    \end{pmatrix}.
\end{equation}
The corresponding structure factors are
\begin{align}
    \gamma_{\bm{q}}^{(1)}
    &=
    \frac{1}{3}
    \left[
        e^{i\bm{q}\cdot\bm{a}_1}
        +
        e^{i\bm{q}\cdot\bm{a}_2}
        +
        e^{-i\bm{q}\cdot(\bm{a}_1+\bm{a}_2)}
    \right],
    \\
    \gamma_{\bm{q}}^{(2)}
    &=
    \frac{1}{3}
    \left[
        \cos\!\left(\bm{q}\cdot(2\bm{a}_1+\bm{a}_2)\right)
        +
        \cos\!\left(\bm{q}\cdot(\bm{a}_1-\bm{a}_2)\right)
        +
        \cos\!\left(\bm{q}\cdot(\bm{a}_1+2\bm{a}_2)\right)
    \right].
\end{align}
The remaining blocks are fixed by the bosonic Bogoliubov structure,
\begin{equation}
    H_{21}^{0}(\bm{q})
    =
    \left[H_{12}^{0}(\bm{q})\right]^{\dagger},
    \qquad
    H_{22}^{0}(\bm{q})
    =
    \left[H_{11}^{0}(-\bm{q})\right]^T .
\end{equation}

The excitation spectrum of the LSWT Hamiltonian in Eq.~\eqref{eq:uud_lswt_ham} contains two quadratic gapless modes at \(\bm{q}=0\), reflecting an accidental degeneracy of the classical UUD state. To lift this accidental degeneracy and open a gap in the spectrum, we include higher-order corrections in the \(1/S\) expansion, which arise from interactions among the H-P bosons:
\begin{equation}
    \mathcal{H}_{\mathrm{int}}
    =
    \mathcal{H}^{(4)}
    +
    \mathcal{O}(S^{-1}),
    \label{eq:ham_int}
\end{equation}
where \(\mathcal{H}^{(4)}\) denotes the quartic terms in the H-P boson operators. Cubic terms are absent because of the collinear nature of the UUD phase and the \(U(1)\) symmetry of the model Hamiltonian~\cite{ChernyshevAL2009}.

To compute the correction to the single-magnon energies, we normal order the quartic interaction \(\mathcal{H}^{(4)}\) in Eq.~\eqref{eq:ham_int}. We distinguish onsite contractions from bond-weighted contractions. The onsite bilinears are defined as
\begin{equation}
    \rho^{\alpha}
    =
    \frac{1}{N_u}
    \sum_{\bm q\in\mathrm{MBZ}}
    \left\langle
    a_{\alpha,\bm q}^{\dagger}
    a_{\alpha,\bm q}
    \right\rangle,
    \qquad
    \delta^{\alpha}
    =
    \frac{1}{N_u}
    \sum_{\bm q\in\mathrm{MBZ}}
    \left\langle
    a_{\alpha,\bm q}
    a_{\alpha,-\bm q}
    \right\rangle .
     \label{eq:boson_bilinears1}
\end{equation}
For contractions associated with nearest-neighbor bonds $(\alpha\neq\beta)$, we define
\begin{align}
    \xi^{\alpha\beta}
    &=
    \frac{1}{N_u}
    \sum_{\bm q\in\mathrm{MBZ}}
    \gamma_{\bm q}^{(1)}
    \left\langle
    a_{\alpha,\bm q}^{\dagger}
    a_{\beta,\bm q}
    \right\rangle,
    \\
    \zeta^{\alpha\beta}
    &=
    \frac{1}{N_u}
    \sum_{\bm q\in\mathrm{MBZ}}
    \gamma_{\bm q}^{(1)}
    \left\langle
    a_{\alpha,\bm q}
    a_{\beta,-\bm q}
    \right\rangle.
    \label{eq:boson_bilinears2}
\end{align}
For same-sublattice second-neighbor bonds $(\alpha=\beta)$, we define analogously
\begin{align}
    \xi_2^{\alpha\alpha}
    &=
    \frac{1}{N_u}
    \sum_{\bm q\in\mathrm{MBZ}}
    \gamma_{\bm q}^{(2)}
    \left\langle
    a_{\alpha,\bm q}^{\dagger}
    a_{\alpha,\bm q}
    \right\rangle,
    \\
    \zeta_2^{\alpha\alpha}
    &=
    \frac{1}{N_u}
    \sum_{\bm q\in\mathrm{MBZ}}
    \gamma_{\bm q}^{(2)}
    \left\langle
    a_{\alpha,\bm q}
    a_{\alpha,-\bm q}
    \right\rangle.
    \label{eq:boson_bilinears3}
\end{align}
$N_u$ is the number of magnetic unit cells.

Normal ordering generates an effective quadratic contribution, which we denote by \(\mathcal{H}_{\mathrm{NO}}\). This correction can be written in the Bogoliubov form
\begin{equation}
    \mathcal{H}_{\mathrm{NO}}
    =
    \frac{1}{2}\sum_{\bm{q}\in\mathrm{MBZ}}
    \begin{pmatrix}
        (\bm{a}_{\bm{q}}^{\dagger})^T & (\bm{a}_{-\bm{q}})^T
    \end{pmatrix}
    H_{\mathrm{NO}}^{(2)}(\bm{q})
    \begin{pmatrix}
        \bm{a}_{\bm{q}} \\
        \bm{a}_{-\bm{q}}^{\dagger}
    \end{pmatrix},
    \qquad
    H_{\mathrm{NO}}^{(2)}(\bm{q})
    \equiv
    \begin{pmatrix}
        H^{\mathrm{NO}}_{11}(\bm{q}) & H^{\mathrm{NO}}_{12}(\bm{q}) \\
        H^{\mathrm{NO}}_{21}(\bm{q}) & H^{\mathrm{NO}}_{22}(\bm{q})
    \end{pmatrix}.
    \label{eq:uud_no_ham}
\end{equation}
The block matrices are given by
\begin{align}
    H_{11}^{\mathrm{NO}}(\bm{q})
    &=
    \begin{pmatrix}
        \mu^A + \mu^A_{J_2}+t^{AA}_{J_2}\gamma_{\bm{q}}^{(2)}
        &
        [t^{AB}]^{*}\gamma_{\bm{q}}^{(1)}
        &
        t^{CA}\gamma_{-\bm{q}}^{(1)}
        \\
        t^{AB}\gamma_{-\bm{q}}^{(1)}
        &
        \mu^B + \mu^B_{J_2}+t^{BB}_{J_2}\gamma_{\bm{q}}^{(2)}
        &
        [t^{BC}]^{*}\gamma_{\bm{q}}^{(1)}
        \\
        [t^{CA}]^{*}\gamma_{\bm{q}}^{(1)}
        &
        t^{BC}\gamma_{-\bm{q}}^{(1)}
        &
        \mu^C + \mu^C_{J_2}+t^{CC}_{J_2}\gamma_{\bm{q}}^{(2)}
    \end{pmatrix},
\end{align}
and
\begin{align}
    H_{12}^{\mathrm{NO}}(\bm{q})
    &=
    \begin{pmatrix}
        \Gamma^A + \Gamma^A_{J_2}+g^{AA}_{J_2}\gamma_{\bm{q}}^{(2)}
        &
        g^{AB}\gamma_{\bm{q}}^{(1)}
        &
        g^{CA}\gamma_{-\bm{q}}^{(1)}
        \\
        g^{AB}\gamma_{-\bm{q}}^{(1)}
        &
        \Gamma^B + \Gamma^B_{J_2}+g^{BB}_{J_2}\gamma_{\bm{q}}^{(2)}
        &
        g^{BC}\gamma_{\bm{q}}^{(1)}
        \\
        g^{CA}\gamma_{\bm{q}}^{(1)}
        &
        g^{BC}\gamma_{-\bm{q}}^{(1)}
        &
        \Gamma^C + \Gamma^C_{J_2}+g^{CC}_{J_2}\gamma_{\bm{q}}^{(2)}
    \end{pmatrix}.
\end{align}
As in the LSWT Hamiltonian, the remaining blocks are fixed by the bosonic Bogoliubov structure,
\begin{equation}
    H_{21}^{\mathrm{NO}}(\bm{q})
    =
    \left[H_{12}^{\mathrm{NO}}(\bm{q})\right]^{\dagger},
    \qquad
    H_{22}^{\mathrm{NO}}(\bm{q})
    =
    \left[H_{11}^{\mathrm{NO}}(-\bm{q})\right]^T .
\end{equation}

The coefficients are defined as
\begin{align}
    \mu^A &= 3J_1\bigl(\rho^B - \rho^C - \operatorname{Re}\xi^{AB} + \operatorname{Re}\zeta^{CA}\bigr), \nonumber\\
    \mu^B &= 3J_1\bigl(\rho^A - \rho^C - \operatorname{Re}\xi^{AB} + \operatorname{Re}\zeta^{BC}\bigr), \nonumber\\
    \mu^C &= 3J_1\bigl(-\rho^A - \rho^B + \operatorname{Re}\zeta^{BC} + \operatorname{Re}\zeta^{CA}\bigr), \nonumber\\
    t^{AB} &= 3J_1\bigl(\xi^{AB} - \tfrac{1}{2}(\rho^A+\rho^B)\bigr), \nonumber\\
    t^{BC} &= 3J_1\bigl(-\xi^{BC} + \tfrac{1}{4}([\delta^B]^*+\delta^C)\bigr), \nonumber\\
    t^{CA} &= 3J_1\bigl(-\xi^{CA} + \tfrac{1}{4}([\delta^C]^*+\delta^A)\bigr), \nonumber\\
    \Gamma^A &= \tfrac{3J_1}{2}\bigl(\xi^{CA} - \zeta^{AB}\bigr), \\
    \Gamma^B &= \tfrac{3J_1}{2}\bigl([\xi^{BC}]^* - \zeta^{AB}\bigr), \nonumber\\
    \Gamma^C &= \tfrac{3J_1}{2}\bigl([\xi^{BC}]^* - \xi^{CA}\bigr), \nonumber\\
    g^{AB} &= 3J_1\bigl(\zeta^{AB} - \tfrac{1}{4}(\delta^A+\delta^B)\bigr), \nonumber\\
    g^{BC} &= 3J_1\bigl(-\zeta^{BC} + \tfrac{1}{2}(\rho^B+\rho^C)\bigr), \nonumber\\
    g^{CA} &= 3J_1\bigl(-\zeta^{CA} + \tfrac{1}{2}(\rho^C+\rho^A)\bigr), \nonumber
\end{align}
and the $J_2$ corrections are
\begin{align}
    \mu^\alpha_{J_2} &= 6J_2\bigl(\rho^\alpha - \operatorname{Re}\xi_2^{\alpha\alpha}\bigr), \\
    t^{\alpha\alpha}_{J_2} &= 6J_2\bigl([\xi_2^{\alpha\alpha}]^* - \rho^\alpha\bigr), \\
    \Gamma^\alpha_{J_2} &= -\tfrac{3J_2}{2}[\zeta_2^{\alpha\alpha}]^*, \\
    g^{\alpha\alpha}_{J_2} &= 3J_2[\zeta_2^{\alpha\alpha} - \delta^\alpha]^*.
\end{align}

At order \(1/S\), the leading correction to the LSWT spectrum is obtained by treating \(H_{\mathrm{NO}}^{(2)}(\bm q)\) perturbatively. We refer to this as the \emph{one-loop correction}. In this treatment, the bilinears in Eqs.~\eqref{eq:boson_bilinears1}--\eqref{eq:boson_bilinears3} are evaluated using the LSWT vacuum. Let \(T_{\bm{q}}\) be the Bogoliubov transformation that diagonalizes the LSWT quadratic Hamiltonian,
\begin{equation}
    H_{0}^{(2)}(\bm{q})
    \xrightarrow[]{T_{\bm{q}}}
    \mathrm{diag}\left(
        \varepsilon_{1,\bm{q}}^{0},
        \varepsilon_{2,\bm{q}}^{0},
        \varepsilon_{3,\bm{q}}^{0},
        -\varepsilon_{1,-\bm{q}}^{0},
        -\varepsilon_{2,-\bm{q}}^{0},
        -\varepsilon_{3,-\bm{q}}^{0}
    \right).
\end{equation}
We write
\begin{equation}
    \begin{pmatrix}
        \bm{a}_{\bm{q}} \\
        \bm{a}_{-\bm{q}}^{\dagger}
    \end{pmatrix}
    =
    T_{\bm{q}}
    \begin{pmatrix}
        \bm{b}_{\bm{q}} \\
        \bm{b}_{-\bm{q}}^{\dagger}
    \end{pmatrix},
    \qquad
    T_{\bm{q}}
    =
    \begin{pmatrix}
        U_{\bm{q}} & V_{\bm{q}} \\
        V_{\bm{q}} & U_{\bm{q}}
    \end{pmatrix}.
\end{equation}
The onsite LSWT contractions are then
\begin{align}
    \rho^{\alpha}
    &=
    \frac{1}{N_u}
    \sum_{\bm q\in\mathrm{MBZ}}
    \sum_{n=1}^{3}
    \left[V_{\bm q}\right]_{\alpha n}^{*}
    \left[V_{\bm q}\right]_{\alpha n},
    \\
    \delta^{\alpha}
    &=
    \frac{1}{N_u}
    \sum_{\bm q\in\mathrm{MBZ}}
    \sum_{n=1}^{3}
    \left[U_{\bm q}\right]_{\alpha n}
    \left[V_{\bm q}\right]_{\alpha n}^{*}.
\end{align}
The bond-weighted LSWT contractions are
\begin{align}
    \xi^{\alpha\beta}
    &=
    \frac{1}{N_u}
    \sum_{\bm q\in\mathrm{MBZ}}
    \gamma_{\bm q}^{(1)}
    \sum_{n=1}^{3}
    \left[V_{\bm q}\right]_{\alpha n}^{*}
    \left[V_{\bm q}\right]_{\beta n},
    \\
    \zeta^{\alpha\beta}
    &=
    \frac{1}{N_u}
    \sum_{\bm q\in\mathrm{MBZ}}
    \gamma_{\bm q}^{(1)}
    \sum_{n=1}^{3}
    \left[U_{\bm q}\right]_{\alpha n}
    \left[V_{\bm q}\right]_{\beta n}^{*},
    \\
    \xi_2^{\alpha\alpha}
    &=
    \frac{1}{N_u}
    \sum_{\bm q\in\mathrm{MBZ}}
    \gamma_{\bm q}^{(2)}
    \sum_{n=1}^{3}
    \left[V_{\bm q}\right]_{\alpha n}^{*}
    \left[V_{\bm q}\right]_{\alpha n},
    \\
    \zeta_2^{\alpha\alpha}
    &=
    \frac{1}{N_u}
    \sum_{\bm q\in\mathrm{MBZ}}
    \gamma_{\bm q}^{(2)}
    \sum_{n=1}^{3}
    \left[U_{\bm q}\right]_{\alpha n}
    \left[V_{\bm q}\right]_{\alpha n}^{*}.
\end{align}
Substituting these LSWT contractions into \(H_{\mathrm{NO}}^{(2)}(\bm{q})\), the one-loop correction to the \(n\)-th single-magnon energy is given by the diagonal matrix element
\begin{equation}
    \delta\varepsilon_{n,\bm{q}}^{\mathrm{1L}}
    =
    \left[
        T_{\bm{q}}^{\dagger}
        H_{\mathrm{NO}}^{(2)}(\bm{q})
        T_{\bm{q}}
    \right]_{nn},
    \qquad
    n=1,2,3.
\end{equation}
The one-loop spectrum is therefore
\begin{equation}
    \varepsilon_{n,\bm{q}}^{\mathrm{1L}}
    =
    \varepsilon_{n,\bm{q}}^{0}
    +
    \delta\varepsilon_{n,\bm{q}}^{\mathrm{1L}}.
\end{equation}
In this perturbative scheme, both \(T_{\bm q}\) and the contractions \(\rho^\alpha\), \(\delta^\alpha\), \(\xi^{\alpha\beta}\), \(\zeta^{\alpha\beta}\), \(\xi_2^{\alpha\alpha}\), and \(\zeta_2^{\alpha\alpha}\) are computed solely from the LSWT Hamiltonian \(H_{0}^{(2)}(\bm q)\).

In the \emph{self-consistent one-loop theory}, by contrast, the normal-ordering correction is fed back into the quadratic Hamiltonian. Namely, one diagonalizes
\begin{equation}
    H_{\mathrm{SCOL}}^{(2)}(\bm q)
    =
    H_{0}^{(2)}(\bm q)
    +
    H_{\mathrm{NO}}^{(2)}(\bm q),
\end{equation}
where \(H_{\mathrm{NO}}^{(2)}(\bm q)\) is evaluated using contractions obtained from the vacuum of \(H_{\mathrm{SCOL}}^{(2)}(\bm q)\) itself. The procedure is iterated according to
\begin{equation}
    \left\{
    \rho^\alpha,
    \delta^\alpha,
    \xi^{\alpha\beta},
    \zeta^{\alpha\beta},
    \xi_2^{\alpha\alpha},
    \zeta_2^{\alpha\alpha}
    \right\}
    \longrightarrow
    H_{\mathrm{NO}}^{(2)}(\bm q)
    \longrightarrow
    H_{\mathrm{SCOL}}^{(2)}(\bm q)
    \longrightarrow
    T_{\bm q}^{\mathrm{SC}}
    \longrightarrow
    \left\{
    \rho^\alpha,
    \delta^\alpha,
    \xi^{\alpha\beta},
    \zeta^{\alpha\beta},
    \xi_2^{\alpha\alpha},
    \zeta_2^{\alpha\alpha}
    \right\},
\end{equation}
until convergence is reached. The resulting excitation spectrum is obtained from the positive Bogoliubov eigenvalues of the converged matrix \(H_{\mathrm{SC}}^{(2)}(\bm q)\). Thus, the one-loop theory treats \(H_{\mathrm{NO}}^{(2)}\) as a leading \(1/S\) perturbative correction evaluated in the LSWT vacuum, whereas the self-consistent theory incorporates \(H_{\mathrm{NO}}^{(2)}\) directly into the quadratic Hamiltonian and updates the bosonic contractions iteratively.

To determine the plateau phase boundaries, we compute the renormalized single-magnon spectrum away from the classical stability field \(B_{\mathrm{cl}}=3J_1S\). In doing so, we assume that the bosonic bilinears are only weakly dependent on the external magnetic field within the plateau. We therefore evaluate these bilinears at \(B_{\mathrm{cl}}\) and keep them fixed when varying \(B\).

For a given field \(B\), the LSWT contribution is obtained from the field-dependent quadratic matrix \(H_{0}^{(2)}(\bm q;B)\), while the normal-ordering correction \(H_{\mathrm{NO}}^{(2)}(\bm q)\) is constructed using the bilinears evaluated at \(B_{\mathrm{cl}}\). In the one-loop approach, the magnon energies are computed by treating \(H_{\mathrm{NO}}^{(2)}(\bm q)\) as the leading \(1/S\) correction to the LSWT spectrum at the corresponding field \(B\). In the self-consistent approach, we instead diagonalize the renormalized quadratic matrix
\begin{equation}
    H_{\mathrm{SCOL}}^{(2)}(\bm q;B)
    =
    H_{0}^{(2)}(\bm q;B)
    +
    H_{\mathrm{NO}}^{(2)}(\bm q),
\end{equation}
where \(H_{\mathrm{NO}}^{(2)}(\bm q)\) is fixed by the self-consistent bilinears obtained at \(B_{\mathrm{cl}}\).

The lower and upper critical fields, \(h_{c1}\) and \(h_{c2}\), are determined by tracking the minimum positive magnon gap as a function of \(B\). The plateau remains stable as long as all renormalized magnon energies are positive. The phase boundaries are identified by the first closing of the gap,
\begin{equation}
    \min_{n,\bm q}\varepsilon_{n,\bm q}(B=h_{c1})=0,
    \qquad
    \min_{n,\bm q}\varepsilon_{n,\bm q}(B=h_{c2})=0,
\end{equation}
on the low-field and high-field sides of the plateau, respectively. Numerically, we locate these gap-closing fields using bisection. In this way, the \(1/S\) correction shifts the plateau boundaries away from the classical value \(B_{\mathrm{cl}}\), thereby producing a finite field window for the magnetization plateau.

\section{High-field analysis of the magnon condensation for $J_2/J_1 > 1/8$} 
To understand the phase diagram of the spin-$1/2$ model right below the saturation field, we use a hard-core boson representation of spin operators, $S^z = 1/2 - a^+ a, S^+ = a, S^- = a^+$, and perform the magnon Bose-Einstein condensation analysis \cite{Nikuni1995,Batista2014}. The analysis begins with the Hamiltonian of magnons in the fully polarized state,
\begin{align}
    H = \sum_{\bf k} \epsilon({\bf k}) a^+_{\bf k} a_{\bf k} + \frac{1}{2N} \sum_{{\bf k}, {\bf k}', {\bf q}} V({\bf q}) a^+_{{\bf k}+{\bf q}} a^+_{{\bf k}'-{\bf q}} a_{{\bf k}'} a_{\bf k},
\end{align}
where the two-particle interaction 
\begin{align}
\label{v_q}
    V({\bf q}) = 2U + 6 J_1 \gamma_1({\bf q}) + 6 J_2 \gamma_2({\bf q})
\end{align}
originates entirely from the $S^z-S^z$ part of the spin exchange interaction and the hard-core constraint term $U \sum_{\bf r} \hat{n}_{\bf r} (\hat{n}_{\bf r}-1)$ which restricts the boson on-site occupation number $\hat{n}_{\bf r}$ to eigenvalues $0$ and $1$ only, for every ${\bf r}$, in the limit $U\to \infty$. Here and below, $\gamma_{1,2}({\bf q})$ are related to \eqref{gammas} via $\gamma_1({\bf q}) = (\gamma_{\bf q} + \gamma_{-{\bf q}})/2 = 1/3 \sum_j \cos({\bf q} \cdot {\bm \delta}_j), \gamma_2({\bf q}) = \gamma^{(2)}_{\bf q} = 1/3 \sum_j \cos({\bf q} \cdot {\bf e}_j)$. The next-nearest neighbor vectors ${\bf e}_j$ of the lattice are related to ${\bf a}_j$ in Eq.~\eqref{gammas} by ${\bf e}_1 = {\bf a}_1, {\bf e}_2 = {\bf a}_3, {\bf e}_3 = {\bf a}_2$.

Magnon dispersion in the fully polarized state is given by 
\begin{equation}
    \epsilon({\bf k}) = 3 J_1 \gamma_1({\bf k})+3 J_2 \gamma_2({\bf k}) + J_1 + J_2 + \tilde{\mu},
\end{equation}
where $\tilde{\mu} = B - B_{\rm sat}$, $B_{\rm sat} = 4(J_1 + J_2)$, measures the distance to the saturation field. The minimum of the dispersion, given by $\tilde\mu$, is reached at the three M-points of the hexagonal Brillouin zone of the triangular lattice: $M_1 = (0, 2\pi/\sqrt{3}), M_2 = (\pi, \pi/\sqrt{3}), M_3 = (-\pi, \pi/\sqrt{3})$. These points are located at the mid-faces of the BZ and, therefore, vectors $2 M_{i=1,2,3}$ are the reciprocal lattice vectors. The presence of the three degenerate minima opens up the competition between the {\em single}-Q and {\em multi}-Q condensation scenarios at $\mu=0$, when magnons become gapless. The single-Q case corresponds to the spontaneous condensation of magnons at {\em one} of the three possible minima $M_i$. The multi-Q case occurs when the magnon experiences coherent condensation at all three $M$ points. 

To describe the condensation, we parameterize the magnon operator in terms of $c$-number condensates $\phi_i$ at momentum $M_i$ \cite{Ye2017}
\begin{align}
    a_{\bf k} = \sqrt{N} \phi_1 \delta_{{\bf k}, M_1} + \sqrt{N} \phi_2 \delta_{{\bf k}, M_2} + \sqrt{N} \phi_3 \delta_{{\bf k}, M_3} + \tilde{a}_{\bf k},
\end{align}
where operator $\tilde{a}_{\bf k}$ describes non-condensed magnons, and use $\bar{\phi}_i$ for the complex conjugate of $\phi_i$.

The competition between the single-Q and multi-Q condensations is captured by the effective Ginzburg-Landau functional 
\begin{align}
\label{GL}
\frac{E}{N}&=\tilde{\mu}(|\phi_{1}|^2+|\phi_{2}|^2+|\phi_{3}|^2)+\frac{1}{2}\Gamma_1(|\phi_{1}|^4+|\phi_{2}|^4+|\phi_{3}|^4)
+\Gamma_2(|\phi_1|^2|\phi_2|^2+|\phi_1|^2|\phi_3|^2+|\phi_2|^2|\phi_3|^2)\notag\\
&+\Gamma_{\rm u}(\bar{\phi}_1^2\phi_2^2+\bar{\phi}_1^2\phi_3^2+\bar{\phi}_2^2\phi_3^2+{\rm h.c.})
\end{align}
where $\Gamma_{1,2,{\rm u}}$ are obtained from the fully renormalized four-point vertex function $\Gamma({\bf k}_1,{\bf k}_2; {\bf q})$.
Namely, 
\begin{align}
    \Gamma_1 = \Gamma(M_1, M_1; {\bf q} = 0); \Gamma_2 = \Gamma(M_1, M_2; {\bf q} = 0) + \Gamma(M_1, M_2; {\bf q} = M_3); 
    \Gamma_{\rm u} = \Gamma(M_1, M_1; {\bf q} = M_3).
\end{align}
A simple analysis of \eqref{GL} shows that $\Gamma_1 < \Gamma_2$ makes it energetically favorable to condense at a single M-point. This situation corresponds to the {\em stripe} phase.
In the opposite case of $\Gamma_1 > \Gamma_2$, the condensate develops coherently at all three M points, resulting in the coplanar ${\bar V}$ state \cite{Ye2017}.

The renormalized vertex function obeys the Bethe-Salpeter equation
\begin{equation}
\label{eq-BS}
    \Gamma({\bf k}_1,{\bf k}_2; {\bf q}) = V({\bf q}) - \frac{1}{N}\sum_{{\bf q}'\in {\rm BZ}}\frac{V({\bf q}-{\bf q}')\Gamma({\bf k}_1,{\bf k}_2; {\bf q}') }{\epsilon({\bf k}_1+{\bf q}')+\epsilon({\bf k}_2-{\bf q}')+\mu}
\end{equation}
where $\mu = 2 \tilde{\mu}$. In the following, we set $J_1 =1$ and $J_2 = j$.

(a) Consider the case ${\bf k}_1 = {\bf k}_2 = M_1$. Then 
\begin{align}
    D_{\rm a}({\bf q}) = \epsilon(M_1+{\bf q})+\epsilon(M_1-{\bf q})+\mu = 2(1+j) + \cos({\bf q}\cdot{\bm \delta}_1) - \cos({\bf q}\cdot{\bm \delta}_2) -\cos({\bf q}\cdot{\bm \delta}_3)\nonumber\\
    - j[\cos({\bf q}\cdot{\bf e}_1) - \cos({\bf q}\cdot{\bf e}_2) + \cos({\bf q}\cdot{\bf e}_3)]+\mu = D_{\rm a}(-{\bf q})
\end{align}
In addition to being even function of ${\bf q}$, the denominator is also invariant under the inversion $(q_x,q_y) \to (-q_x,q_y)$, which results in invariance under ${\bm \delta}_2 \leftrightarrow {\bm \delta}_3$ and ${\bf e}_1 \leftrightarrow {\bf e}_3$. The interaction potential $V({\bf q})$ shares these properties. This suggests the following ansatz for the vertex function 
\begin{align}
    \Gamma({\bf k}_1,{\bf k}_2; {\bf q}) = \sum_{i=0}^4 \eta_i \psi_i({\bf q})
\end{align}
with the basis vectors 
\begin{align}
\label{psi-basis}
    \psi_0({\bf q}) = 1; \psi_1({\bf q}) = \sqrt{2} \cos({\bf q}\cdot{\bm \delta}_1); \psi_2({\bf q}) = \cos({\bf q}\cdot{\bm \delta}_2) + \cos({\bf q}\cdot{\bm \delta}_3); \nonumber\\ \psi_3({\bf q}) = \cos({\bf q}\cdot{\bf e}_1) + \cos({\bf q}\cdot{\bf e}_3); \psi_4({\bf q}) = \sqrt{2} \cos({\bf q}\cdot{\bf e}_2).
\end{align}
These functions form an orthonormal basis, $1/N \sum_{{\bf q} \in {\rm BZ}} \psi_i({\bf q}) \psi_j({\bf q}) = \delta_{ij}$. Not surprisingly, the interaction potential \eqref{v_q} can be written in this basis as well, $V({\bf q}) = \sum_{i=0}^4 v_i \psi_i({\bf q})$, where $v_i = (2U, \sqrt{2}, 2, 2j, \sqrt{2}j)$.

From here, we reduce \eqref{eq-BS} to the matrix equation by multiplying its both sides with $\psi_j({\bf q})$ and integrating over ${\bf q}$, for $j=0...4$. Since $\psi_0=1$, the first of these equations is obtained by integrating both sides of \eqref{eq-BS} over ${\bf q}$ and using $1/N \sum_{\bf q} \psi_n({\bf q}) = 0$ for $n=1,2,3,4$. We obtain $\eta_0/(2U) = 1 - 1/N \sum_{{\bf q}'} \Gamma({\bf k}_1,{\bf k}_2; {\bf q}')/D_{\rm a}({\bf q}')$ and note that in the hard-core $U\to \infty$ limit the left-hand side vanishes. Upon introducing 
\begin{align}
\label{taus}
    \tau_{ij}=\tau_{ji} = \frac{1}{N} \sum_{{\bf q} \in {\rm BZ}} \frac{\psi_i({\bf q}) \psi_j({\bf q})}{D_{\rm a}({\bf q})},
\end{align}
this equation can be compactly written as $\sum_{i=0}^4 \tau_{0i} \eta_i = 1$.
Next, multiply \eqref{eq-BS} by $\psi_{j=1,2,3,4}$ and integrate over ${\bf q}$. Using easily derived identities 
\begin{eqnarray}
    &&\frac{1}{N} \sum_{{\bf q} \in {\rm BZ}} \psi_{j=1,4}({\bf q}) \psi_{i=1,4}({\bf q} - {\bf q}') = \frac{1}{\sqrt{2}} \psi_{j=1,4}({\bf q}') \delta_{ji}, \nonumber\\
    &&\frac{1}{N} \sum_{{\bf q} \in {\rm BZ}} \psi_{j=2,3}({\bf q}) \psi_{i=2,3}({\bf q} - {\bf q}') = \frac{1}{2} \psi_{j=2,3}({\bf q}') \delta_{ji}
\end{eqnarray}
we obtain
\begin{align}
    \eta_j + \frac{1}{N} \sum_{{\bf q}' \in {\rm BZ}} \sum_{i=0}^4 \eta_i \frac{\psi_i({\bf q}')}{D_{\rm a}({\bf q}')}\Big[\frac{v_1}{\sqrt{2}} \psi_1({\bf q}') \delta_{j1} + \frac{v_2}{2} \psi_2({\bf q}') \delta_{j2} + \frac{v_3}{2} \psi_3({\bf q}') \delta_{j3} + \frac{v_4}{\sqrt{2}} \psi_4({\bf q}') \delta_{j4}\Big] = v_j
\end{align}
Altogether, \eqref{eq-BS} turns into the matrix equation for vector $\vec{\eta} = (\eta_0,\eta_1,\eta_2,\eta_3,\eta_4)^{\rm t}$
\begin{align}
\label{Gamma1}
\begin{pmatrix}
\tau_{00}&\tau_{01}&\tau_{02}&\tau_{03}&\tau_{04}\\
\tau_{10}&1+\tau_{11}&\tau_{12}&\tau_{13}&\tau_{14}\\
\tau_{20}&\tau_{12}&1+\tau_{22}&\tau_{23}&\tau_{24}\\
j \tau_{30}&j \tau_{13}&j \tau_{23}&1+j\tau_{33}& j \tau_{34}\\
j \tau_{40}&j \tau_{14}&j \tau_{24}&j \tau_{34}&1+j \tau_{44}
\end{pmatrix}\begin{pmatrix}
\eta_0\\\eta_1\\ \eta_2\\ \eta_3\\ \eta_4
\end{pmatrix}=\begin{pmatrix}
1\\\sqrt{2}\\2\\2j\\\sqrt{2}j
\end{pmatrix}
\end{align}
Once $\vec{\eta}$ is found, $\Gamma_{1,{\rm u}}$ follow via
\begin{align}
    \Gamma_1 = \sum_{i=0}^4 \eta_i \psi_i({\bf q}=0) = \vec{\psi}({\bf q}=0) \cdot \vec{\eta}; 
    \,\, \Gamma_{\rm u} = \sum_{i=0}^4 \eta_i \psi_i({\bf q}=M_3) = \vec{\psi}({\bf q}=M_3) \cdot \vec{\eta}.
\end{align}
Here $\vec{\psi} = (\psi_0,\psi_1,\psi_2,\psi_3,\psi_4)$ is the vector of the basis functions \eqref{psi-basis}, and we denote the vector on the right-hand side as $\vec{w} = (1, \sqrt{2}, 2, 2j, \sqrt{2}j).$

Each of $\tau_{ij}$ diverges logarithmically in the limit $\mu\to 0$. The divergence comes from the neighborhood of points ${\bf q}' =0, M_2, M_3$ in $D_{\rm a}({\bf q}')$. The diverging contribution to $\tau_{ij}$ is found to be 
\begin{align}
\label{taus-log}
    \tilde{\tau}_{ij} = (\psi_i(0) \psi_j(0) + 2 \psi_i(M_3)\psi_j(M_3)) L_{\rm a}(j,\mu), 
\end{align}
where 
\begin{align}
    L_{\rm a}(j,\mu) = \frac{1}{4\pi \sqrt{4j - (3j-1)^2}}\ln(\frac{1}{\mu}).
\end{align}
We then split $\tau_{ij} = \tilde{\tau}_{ij} + \tau^0_{ij}$, where $\tau^0_{ij} = \tau_{ij} - \tilde{\tau}_{ij}$ is the regular contribution which is evaluated numerically by calculating the difference of the integral $\tau_{ij}$ \eqref{taus} and \eqref{taus-log} in the $\mu \to 0$ limit. Practically, we find $\tau^0_{ij}$ to reach a $\mu$-independent value for $\mu \sim 10^{-6} - 10^{-8}$.

Finally, we solve \eqref{Gamma1} for $\vec{\eta}$ by inverting the matrix with the help of Mathematica and expanding the result in inverse powers of diverging $L_{\rm a}(j,\mu\to 0)$ to find
\begin{align}
\label{casea}
    \Gamma_1(j,\mu) = \frac{1}{L_{\rm a}(j,\mu)} + \frac{c_{\rm a}(j)}{L^2_{\rm a}(j,\mu)} + ...; 
    ~~~\Gamma_{\rm u}(j,\mu) = \frac{c_{\rm u}(j)}{L^2_{\rm a}(j,\mu)} + ...
\end{align}
Here, $j$-dependent coefficients $c_{\rm a/u}(j)$ are determined by the {\em regular} $\tau^0_{ij}$ contributions. We find $c_{\rm a}(1/3) \approx -0.085$ and $c_{\rm u}(1/3) \approx -0.011$. The fact that $\Gamma_{\rm u} \ll \Gamma_1$ is consistent with the previous large-$S$ calculation \cite{Ye2017}.

(b) Now consider ${\bf k}_1 = M_1, {\bf k}_2 = M_2$. Then 
\begin{align}
    D_{\rm b}({\bf q}) = \epsilon(M_1+{\bf q})+\epsilon(M_2-{\bf q})+\mu = 2(1+j) -2  \cos({\bf q}\cdot{\bm \delta}_3) 
    - 2j\cos({\bf q}\cdot{\bf e}_1) +\mu = D_{\rm b}(-{\bf q})
\end{align}
This suggests a slightly different set of basis functions
\begin{align}
    \chi_0({\bf q}) = 1; \chi_1({\bf q}) = \cos({\bf q}\cdot{\bm \delta}_1) + \cos({\bf q}\cdot{\bm \delta}_2); \chi_2({\bf q}) = \sqrt{2} \cos({\bf q}\cdot{\bm \delta}_3) \nonumber\\ 
    \chi_3({\bf q}) = \sqrt{2} \cos({\bf q}\cdot{\bf e}_1); \chi_4({\bf q}) = \cos({\bf q}\cdot{\bf e}_2) + \cos({\bf q}\cdot{\bf e}_3).
\end{align}
We expand $\Gamma({\bf k}_1,{\bf k}_2; {\bf q}) = \sum_{i=0}^4 \eta'_i \chi_i({\bf q})$ and $V({\bf q}) = \sum_{i=0}^4 v'_i \chi_i({\bf q})$, where $v'_i = (2U, 2, \sqrt{2}, \sqrt{2}j, 2j)$. It is straightforward to derive the matrix equation for $\vec{\eta'}$, which is formulated in terms of 
\begin{align}
    s_{ij} = \frac{1}{N} \sum_{{\bf q} \in {\rm BZ}} \frac{\chi_i({\bf q}) \chi_j({\bf q})}{D_{\rm b}({\bf q})},
\end{align}
and has vector $\vec{w'}=(1, 2, \sqrt{2}, \sqrt{2} j, 2j)$ in the right-hand side. Solving it gives
\begin{align}
    \Gamma_2 = \sum_{i=0}^4 \eta'_i (\chi_i({\bf q} =0) + \chi_i({\bf q} =M_3)).
\end{align}
In distinction from the previous case, here $s_{ij}$ diverges logarithmically when ${\bf q}' =0 \, {\rm or} \, M_3$. The diverging contribution $\tilde{s}_{ij}$ reads 
\begin{align}
\label{sus-log}
    \tilde{s}_{ij} = (\chi_i(0) \chi_j(0) +  \chi_i(M_3)\chi_j(M_3)) L_{\rm b}(j,\mu), \, {\rm where} \, L_{\rm b}(j,\mu) = \frac{1}{4\pi \sqrt{4j}}\ln(\frac{1}{\mu}).
\end{align}
At the end of the day, we find that in the $\mu\to 0$ limit 
\begin{align}
\label{caseb}
    \Gamma_2(j,\mu) = \frac{1}{L_{\rm b}(j,\mu)} + \frac{c_{\rm b}(j)}{L^2_{\rm b}(j,\mu)} + ... \, {\rm with} \, c_{\rm b}(1/3) \approx -0.092.
\end{align}
Therefore, 
\begin{align}
    \Gamma_1(j,\mu) - \Gamma_2(j,\mu) = 4\pi \frac{\sqrt{4j-(3j-1)^2} - \sqrt{4j}}{\ln(1/\mu)} + O(\frac{1}{\ln^2(1/\mu)})
\end{align}
That is, for every $j\neq 1/3$, $\Gamma_1 < \Gamma_2$ because the coefficient of the leading, $1/\ln(1/\mu)$, contribution is negative. The subleading $1/\ln^2(1/\mu)$ terms can not change this outcome in the  $\mu\to 0$ limit. This means the {\em stripe} magnetic order.

The only exception to this occurs at $j=1/3$, when the leading contributions cancel out. At that point, $1/\ln^2(1/\mu)$ terms `flip' the outcome and lead to $\Gamma_1 > \Gamma_2$. At this special point, it is the $\bar{V}$ phase that is stabilized just below the saturation field. Our result, {\em exact} for the spin $S=1/2$ $J_1-J_2$ triangular lattice antiferromagnet, confirms previous semi-classical, $S \gg 1$, analysis where the direct transition from the polarized to ${\bar V}$ state at $J_2 = J_1/3$ was first suggested \cite{Ye2017}.

\end{document}